\documentclass[onecolumn]{pasj00}

\begin{document}
\SetRunningHead{Iwata et al.}{Lyman Break Galaxies at $z \sim 5$}
\Received{2002/08/25}
\Accepted{2003/01/06}

\title{Lyman Break Galaxies at $z \sim 5$: Luminosity Function\thanks{
Based on data collected at Subaru Telescope, 
which is operated by the National Astronomical Observatory of Japan.}}

\author{Ikuru \textsc{Iwata},\altaffilmark{1}
Kouji \textsc{Ohta},\altaffilmark{1}
Naoyuki \textsc{Tamura},\altaffilmark{1,2}\\
Masataka \textsc{Ando},\altaffilmark{1}
Shinpei \textsc{Wada},\altaffilmark{1} 
Chisato \textsc{Watanabe},\altaffilmark{1}\\
Masayuki \textsc{Akiyama},\altaffilmark{3}
and Kentaro \textsc{Aoki}\altaffilmark{3,4}
}%
\altaffiltext{1}{Department of Astronomy, Faculty of Science, 
Kyoto University, \\
Sakyo-ku, Kyoto 606-8502} 
\altaffiltext{2}{Department of Physics, University of Durham,  
Durham DH1 3LE UK}
\altaffiltext{3}{Subaru Telescope, National Astronomical Observatory of
Japan,\\ 650 North A'ohoku Place, Hilo, Hawaii 96720 U.S.A}
\altaffiltext{4}{Institute of Astronomy, University of Tokyo,\\
2-21-1 Osawa, Mitaka, Tokyo 181-0015}
\email{iwata@kusastro.kyoto-u.ac.jp}

\KeyWords{galaxies: evolution --- galaxies: statistics -- 
} 

\maketitle

\begin{abstract}
We present results of a search for Lyman break galaxies (LBGs) at 
$z \sim 5$ in a 618 square-arcmin field including the Hubble Deep 
Field-North (HDF-N) taken by Subaru Prime Focus Camera.
Utilizing the published redshift data of the HDF-N and its 
flanking fields, the color selection criteria are chosen so that 
LBGs are picked out most efficiently and least 
contaminated by foreground objects.
The numbers of LBG candidates detected 
are 96 in 23.0 $\leq$ $I_c$ mag $\leq$ 24.5, 
and 310 in $23.0 \leq I_c \leq 25.5$. 
There is a hint of the deficiency of bright blue galaxies, 
although it is not as clear as has been 
suggested for LBGs at $z \sim$ 3 to 4. 
With 305 LBG candidates in a 575 square-arcmin field, 
the rest-frame UV luminosity function of LBGs at 
$4.4 \lesssim z \lesssim 5.3$ is derived statistically, 
by considering both contamination by objects 
at the intermediate redshift and incompleteness of the survey.
The fraction of contamination is estimated to be 
$\sim 50 \%$ in the faintest magnitude range.
The completeness of the survey is $\sim$80\% at the bright part 
of the sample, and $\sim20$\% in the faintest magnitude 
range ($25.0 < I_c \leq 25.5$).
The luminosity function of LBG candidates at $z \sim$ 5 
does not show a significant difference from those of LBGs at 
$z \sim$ 3 and 4, though there might be a slight decrease in the 
fainter part.
The UV luminosity density within the observational limit 
is 0.56 -- 0.69 times smaller than that obtained for LBGs at $z \sim 3$, 
depending on the adopted cosmology and the integration 
range of the luminosity function.
This decrease in the UV luminosity at $z \sim 5$ compared to 
that at $z \sim 3$ is due to the smaller number density of faint 
galaxies at $z \sim 5$.
The similarity of the luminosity functions at redshifts 5 to 3 
implies that most of the LBGs at $z \sim 5$ should have faded out 
at $z \sim 3$ and the LBGs at $z \sim 5$ are different galaxies 
from those seen at $z \sim 3$, 
if we take face values for ages of the LBGs at $z \sim 3$ obtained 
by the SED fitting in which a continuous star formation in an 
individual galaxy is assumed.
However, if the star formation in LBGs is sporadic, the similarity 
of the luminosity function at $z \sim 3$ and 5 would be explained. 
Such sporadic star formation has been suggested by hydrodynamical 
simulations and semi-analytic models with collisional starbursts, 
and the trend of the cosmic star formation history 
predicted by these studies resembles to 
that estimated from an integration of the UV luminosity 
density within the observational limit.
\end{abstract}

\section{Introduction}

In the past several years, observational 
studies of galaxy formation and evolution
have progressed remarkably.
One of the most important progresses is that statistical nature of
high redshift galaxies has been revealed through a finding of many
normal (non-AGN) galaxies at high redshift, which
has been enabled by the broadband color selection method pioneered 
by Steidel and his collaborators (e.g., \cite{S92}; Steidel et al. 1996a).
Since the method employs only broad
band imaging, it does not need a huge amount of telescope time
such as optical spectroscopy of faint galaxies,
and it efficiently finds galaxies at a targeted redshift.
In fact, the redshifts of more than 80\% of the galaxies selected by this
method (called Lyman break galaxies) are confirmed by 
optical spectroscopy (Steidel et al. 1996a; \cite{S99}).

Since their first discovery in the late 1990s, the Lyman break galaxies
(LBGs) at $z \sim3$ have been extensively investigated in various ways 
and knowledge on their nature has been accumulated.
Their rest UV spectra resemble those of nearby active star-forming
galaxies (Steidel et al. 1996b). 
On-going star formation rates estimated from UV luminosities and
from emission lines in rest-frame optical band are up to several tens of
$\MO$ yr$^{-1}$ (e.g., Steidel et al. 1996b; \cite{Pettini(2001)}). 
Their stellar population has been investigated by comparing the 
optical to near-infrared spectral energy distributions (SEDs) 
with stellar population synthesis models
(\cite{Sawicki(1998)}; \cite{Papovich(2001)}; \cite{Shapley(2001)}).
It is found that most of the LBGs at $z \sim 3$ are well fitted by models
dominated by young stellar population with ages
ranging from 10 Myr to 1 Gyr with moderate dust extinction 
($E(B-V) \sim$ 0.1 -- 0.4) and their typical stellar mass is 
estimated to be $\sim$ $10^{10}$ $\MO$, similar to a typical bulge 
mass of a present-day giant galaxy.
Metal abundance in ionized gas is estimated through rest optical 
emission line ratios to be sub-solar similar to those of
Large and Small Magellanic Clouds 
(\cite{Teplitz(2000)}; \cite{Pettini(2001)}).
The clear signal of spatial clustering of the LBGs at $z \sim 3$ has
been reported (\cite{S98}; \cite{Giavalisco(1998)}), and 
the comparison with
the models of cold dark matter (CDM) halos suggests that their clustering
properties of the LBGs can be ascribed to the 
biased galaxy formation \citep{Giavalisco(2001)}.

\citet{S99} extend their ground-based LBG search to $z \sim 4$.
They surveyed a 828 arcmin$^2$ area in total which is 
comprised of 10 fields
and found 244 galaxies with $I_{\mathrm{AB}} \leq 25.0$ mag.
The luminosity function of the LBGs at $z \sim 4$
remarkably resembles that at $z \sim 3$,
indicating that no significant evolution of
luminosity function occurs from $z=4$ to $z=3$ (about 0.6 Gyr 
for the cosmology with $\Omega_{\mathrm{M}}=0.3$, $\Omega_\Lambda=0.7$ 
and $H_0 = 65$ km s$^{-1}$ Mpc$^{-1}$).
\citet{Ouchi01a} also report the result of their own LBG survey for $z
\sim 4$ in the field of {\it SUBARU/XMM} Deep Survey. They found
that the luminosity function almost agrees with that obtained
by \citet{S99} and that the significant clustering is also seen.

The LBGs make the statistical sample of 
UV luminous objects at high redshift.
Since the UV luminosity density is considered to be 
correlated with star formation rate per unit volume, 
the LBGs have opened up a possibility to examine the star formation 
history in the early universe (Madau et al. 1996, 1998).
\citet{S99} found that the UV luminosity density at $z \sim 4$ 
is almost comparable to that at $z \sim 3$.
If the correction for the dust extinction is applied, 
the star formation densities at $z=3$ and 4 are comparable to
the value at $z=1-2$.
The similarity of the luminosity function of LBGs at $z=3$ and at $z=4$ 
and the rather flat (extinction corrected) star formation density
history in $1 < z < 4$ are quite interesting and may be a key 
property to understand star formation history of galaxies.
As mentioned above, the ages of the LBGs at $z=3$ 
are estimated to be 10 Myr to 1 Gyr with a median value of
0.1--0.3 Gyr based on the SED fitting analyses by assuming 
a continuous star formation.
If these ages can be regarded as time intervals from the
onset of the first star formation, the LBGs at $z\sim 3$ should be
very young and most of them should form at $z<4$
(e.g., \cite{Papovich(2001)}).
Accordingly, the star formation density contributed by the LBGs at
$z\sim3$ should be lower by a factor of about 10 at $z>4$ 
\citep{Ferguson(2002)} and
the LBGs at $z \sim 3$ must be different from 
the LBGs at $z \gtrsim 4$.

By using semi-analytic models based on the hierarchical clustering
scenario, several authors have tested 
whether the observed number densities 
and star formation densities at $z \sim 3$ and $z \sim 4$ 
can be reproduced (e.g., \cite{Baugh(1998)}; \cite{Somerville(2001)}; 
\cite{Balland(2002)}).
\citet{Somerville(2001)} examine the nature of these LBGs
using semi-analytic models including starburst induced 
by collision of galaxies as well as continuous star formation.
They claim that their model with collisional starburst 
reproduces the observed luminosity functions of LBGs
both at $z=3$ and $z=4$ with a reasonable amount of dust extinction, 
and the estimated star formation density is claimed to be 
compatible with observations. On the other hand, the model 
with constant star formation efficiency (which does not include 
collisional starburst) fails to reproduce the observed luminosity 
function of LBGs at $z \sim 3$, and star formation density 
declines rapidly at $z > 2$.

Pushing these studies back to the earlier universe and
disclosing the nature of higher redshift galaxies 
is an obvious next target.
The extension of the observational data toward higher redshift 
provides a stronger constraint on physical 
processes of the galaxy evolution through the comparison 
with theoretical models.
It is also indispensable to study the epoch of the generation 
of the first galaxies.
The study at the higher redshifts has a further importance
because it is approaching to a suggested epoch of the cosmological
reionization ($z \gtrsim 6$; \cite{Becker(2001)}). 
The number densities of the LBGs may provide a constraint on
determining the contribution of the star forming galaxies to the 
ionizing background radiation \citep{Madau(1999)}.
The discoveries of several star forming galaxies at $z \gtrsim 5$ 
have been already reported 
(e.g., \cite{Dey(1998)}; \cite{Weymann(1998)};
\cite{Spinrad(1998)}; \cite{Dawson(2001)}; \cite{Ellis(2001)};
\cite{Dawson(2002)}).
Their properties are important in the context of the investigation
of the nature of galaxies at high redshift. However, 
since there has not been a volume-limited large sample,
statistical properties of galaxies at 
$z \sim 5$ remain unknown.
Finding many galaxies at $z\sim5$ systematically and studying their
statistical nature is necessary for the first step.
We should carry out a survey with a sufficiently large search
area at a deep limiting magnitude.
However, it is not easy in technical aspects.
The distance to the objects are getting larger and thus apparent
magnitude and surface brightness are getting fainter.
Consequently a number density at an apparent magnitude decreases
rapidly.
In addition, the observing band shifts to the redder wavelength
region where OH sky emission increases the background noise.
Therefore a large aperture of a telescope and a wide field of
view are required to perform this task.

Here we present the first results of our survey of LBGs at $z \sim 5$
using the Subaru Prime Focus Camera (Suprime-Cam;
\cite{Miyazaki(2002)}), 
utilizing its unique capability of a wide field of 
view among 8m class telescopes.
We choose the sky area including the Hubble Deep Field-North 
(HDF-N; \cite{Williams(1996)}),
because there is plenty of information of 
galaxies with a wide redshift range ($0 < z < 6$).
This database is quite powerful to examine our method for 
searching for galaxies at $z \sim 5$  and to obtain 
a reliable sample of candidates, under the circumstances
that considerably time consuming optical follow-up spectroscopy
must be carried out after making a candidate catalog.
The field is also the sky area where many deep surveys in
variety of wavelengths such as X-ray, infrared and radio.
We will be able to utilize these data to investigate nature of
LBG population in other wavebands.

This paper is organized as follows. 
In section 2, the observations and procedures of data reduction are
described.
In section 3, criteria for finding LBGs at $z \sim 5$ and 
the number of objects detected are presented. 
Properties of the LBG candidates are also described. 
Then in section 4, 
we quantify the corrections needed to apply to the observed data 
for deriving the luminosity function statistically, 
and the resultant luminosity 
function of LBGs at $z\sim 5$ is presented. 
In section 5, implications of our findings to the 
evolution process of LBGs
and the evolution of the UV luminosity density over the 
cosmic time scale is discussed.
The contribution of LBGs at $z \sim 5$ 
to the UV ionizing field is also mentioned. 
In section 6 we give a summary and conclusion of our findings.

We use a set of cosmological parameters of
$H_0 = 65$ km s$^{-1}$ Mpc$^{-1}$, $\Omega_\mathrm{M} = 0.3$, and 
$\Omega_\Lambda =0.7$ as a
primary set, and in some cases we also use an Einstein-de Sitter 
universe model 
with $H_0 = 50$ km s$^{-1}$ Mpc$^{-1}$, $\Omega_\mathrm{M} = 1.0$ and 
$\Omega_\Lambda = 0$ 
which had been commonly used as a standard set of cosmological 
parameters.
The magnitude system is based on Vega magnitude
except for the cases otherwise noted.

\section{Observation and Data Reduction}

\subsection{Filter Selection}

\begin{figure}
  \begin{center}
    \FigureFile(80mm,50mm){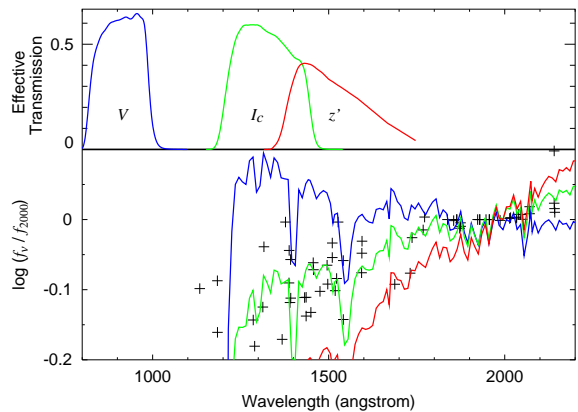}
  \end{center}
  \caption{Model UV spectra of LBGs at $z\sim5$, which are used for the filter
 selection and for the evaluation of the incompleteness of our survey
 via Monte Carlo simulation (in section 5.2). 
 The bluest (extinction-free) spectrum is shown in the blue solid line.
 The green and red lines are spectra of the redder
 LBGs attenuated by dust with $E(B-V)=$ 0.2 and 0.4, respectively.
The attenuation by the intergalactic matter is taken into
account (Madau et al. 1996).
 Points are photometric data of the LBGs in the HDF-N at $z \sim 3$ 
 \citep{Papovich(2001)}; we intend to cover the various SEDs at 
 $z \sim 3$ by model spectra.
 All flux densities are normalized at 2000 \AA. 
 In the upper side of the figure we show the 
 effective band-passes of the Suprime-Cam with $V$, $I_c$ and $z'$
 filters, which are blueshifted as they work for an object at
 $z=5$.
}
 \label{fig_sed}
\end{figure}

Prior to the observation, the combination of the filters suited to
isolate the $z \sim 5$ population was explored using mock
galaxy spectra constructed with stellar population synthesis codes.
Although it is not obvious that SEDs of LBGs at 
$z \sim 5$ are similar to those at $z \sim 3$, 
we adopted model SEDs of LBGs at $z \sim 5$ to 
enclose various rest-frame UV SEDs of LBGs at $z \sim 3$.
The model spectrum for young, actively star-forming galaxies, by which 
we intend to model the SED of LBGs with the bluest UV color, 
was constructed using the stellar synthesis code developed by 
\citet{Kodama(1997)}.
We included the attenuation due to the inter-galactic medium
following the prescription by \citet{Madau(1995)}. 
In order to have the model spectra delineate the
redder SED of LBGs, we adopted to take into account the
reddening effect by dust within the object itself
using the extinction curve given by \citet{Calzetti(2000)}
with $E(B-V)$ from 0.0 to 0.4.
In figure \ref{fig_sed} we show the SEDs of the model spectra with 
$E(B-V)$ = 0.0, 0.2 and 0.4, as well as the rest-frame photometric data
points of LBGs at $z \sim 3$ in the HDF-N \citep{Papovich(2001)}.
The model spectra reasonably cover the UV colors of the LBGs at 
$z \sim 3$.
The adoption of the dust extinction is rather arbitrary to
reproduce the redder colors of LBGs.
The interpretation is, however, often used to derive
color excesses in LBGs (e.g., \cite{S99}).

We also prepared model spectra of spiral and elliptical galaxies for 
examining the efficiency of eliminating galaxies at
intermediate redshifts.
As spectra of spiral galaxies we used those of Sbc and Scd
galaxies derived by \citet{Coleman(1980)}.
As a spectrum of an elliptical galaxy we employed the model
spectrum by \citet{Kodama(1997)} which reproduces the color-magnitude
relation of elliptical galaxies observed in the local universe.
Then we calculated observed spectra for each type of galaxy from 
$z=0$ to $z=6$ with a step of 0.1. 
We did not consider any effect of evolution for
all types of galaxies.

\begin{figure}
  \begin{center}
    \FigureFile(70mm,48mm){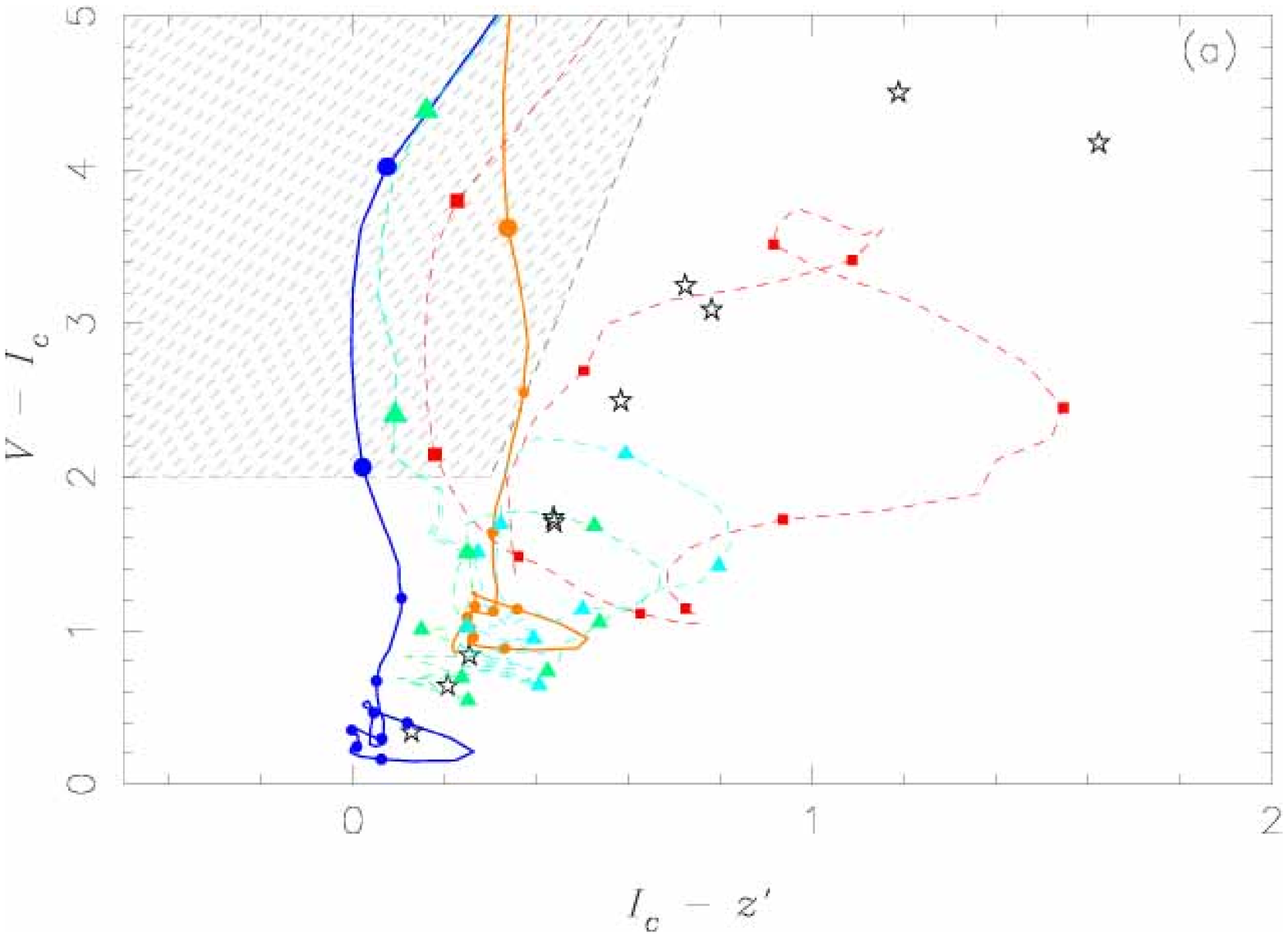}
    \FigureFile(70mm,48mm){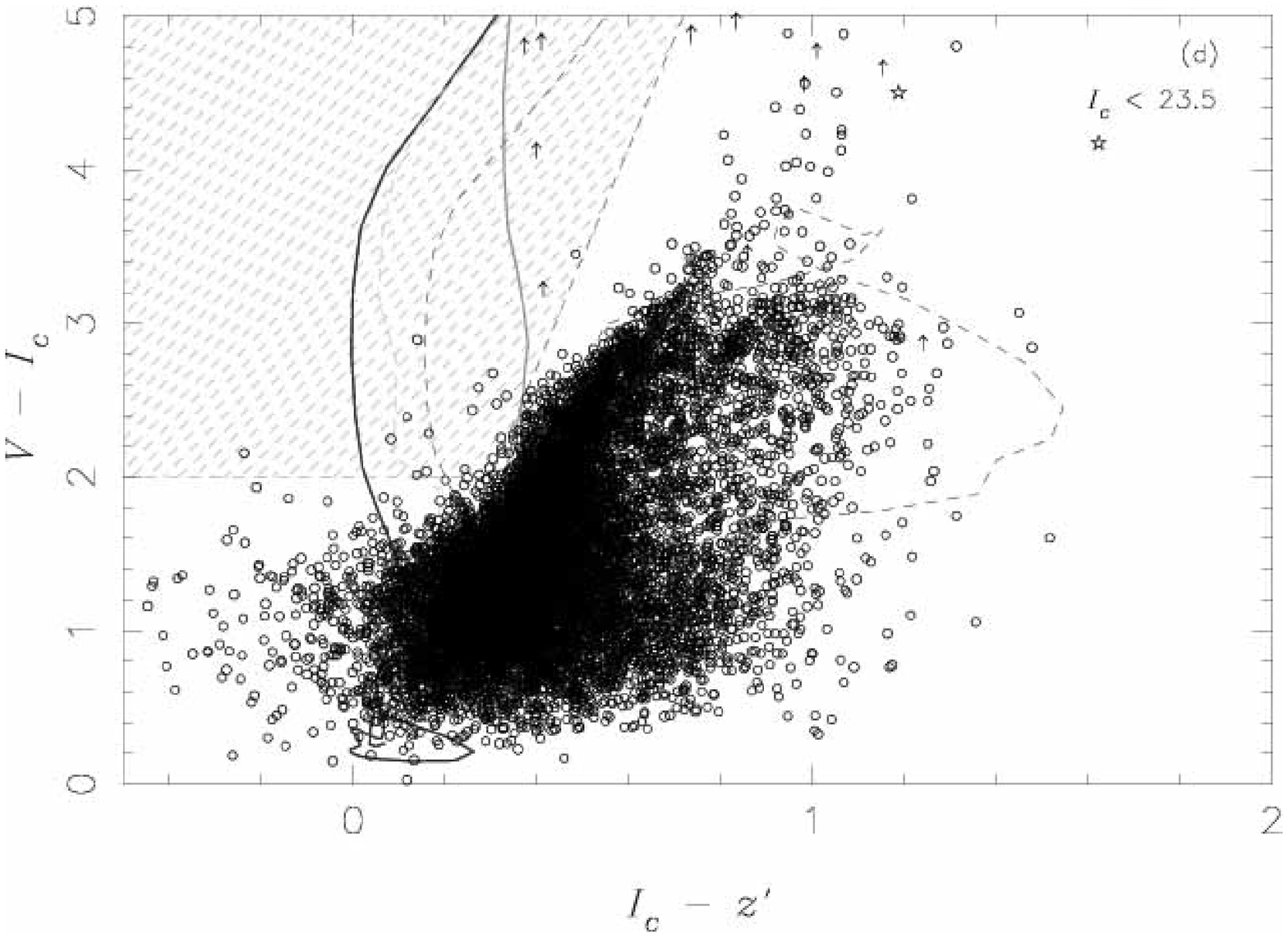}\\
    \FigureFile(70mm,48mm){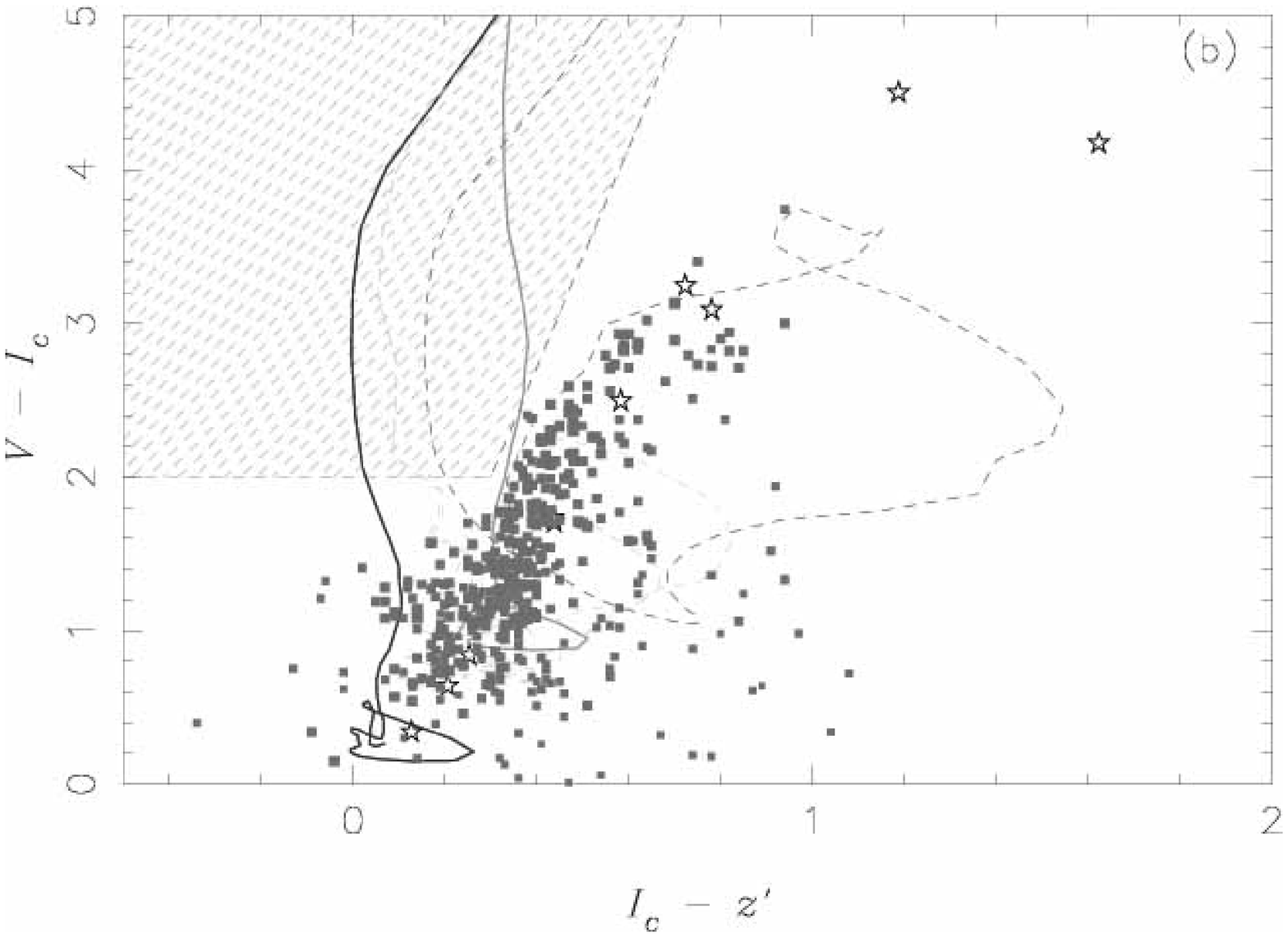}
    \FigureFile(70mm,48mm){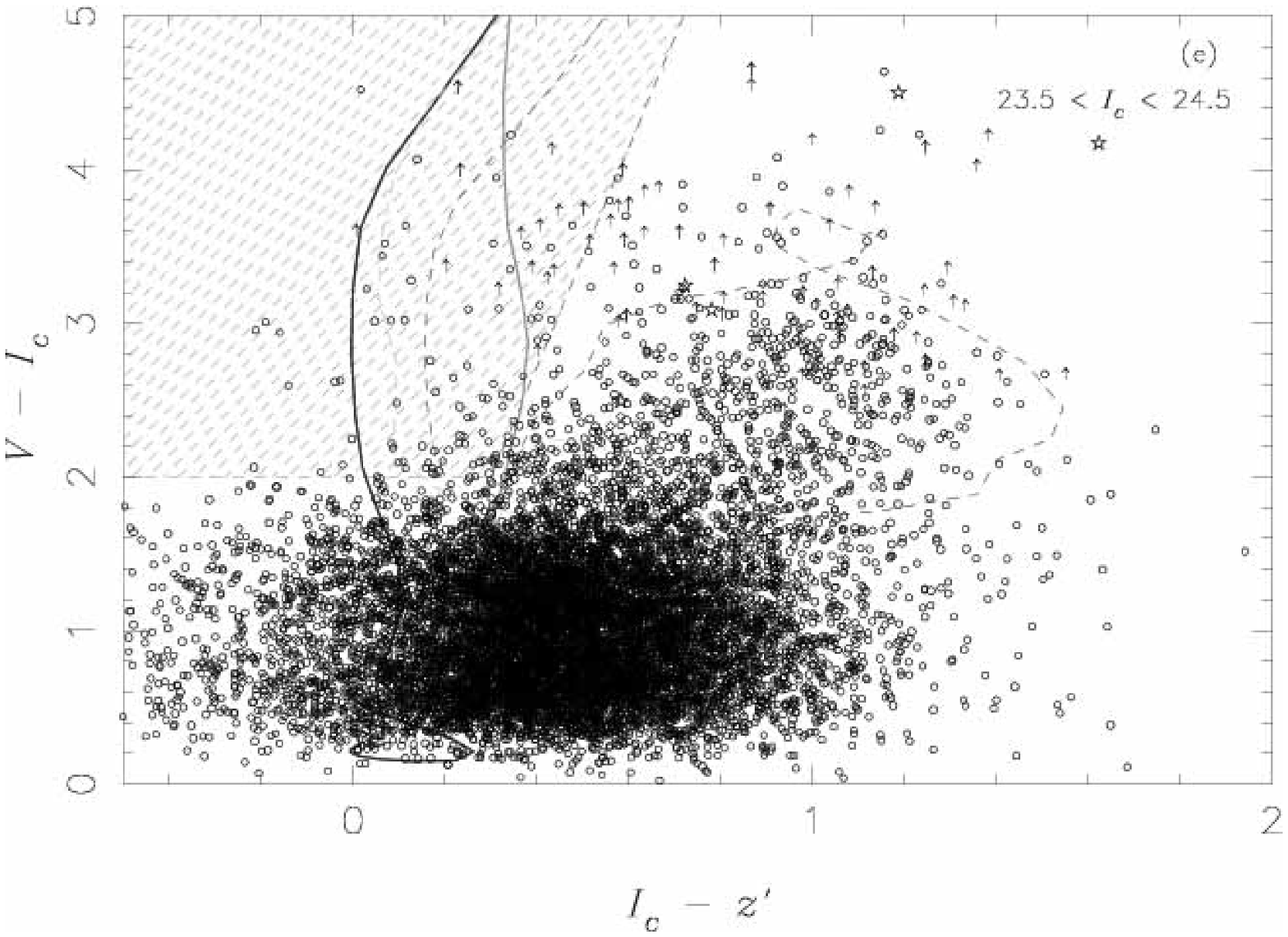}\\
    \FigureFile(70mm,48mm){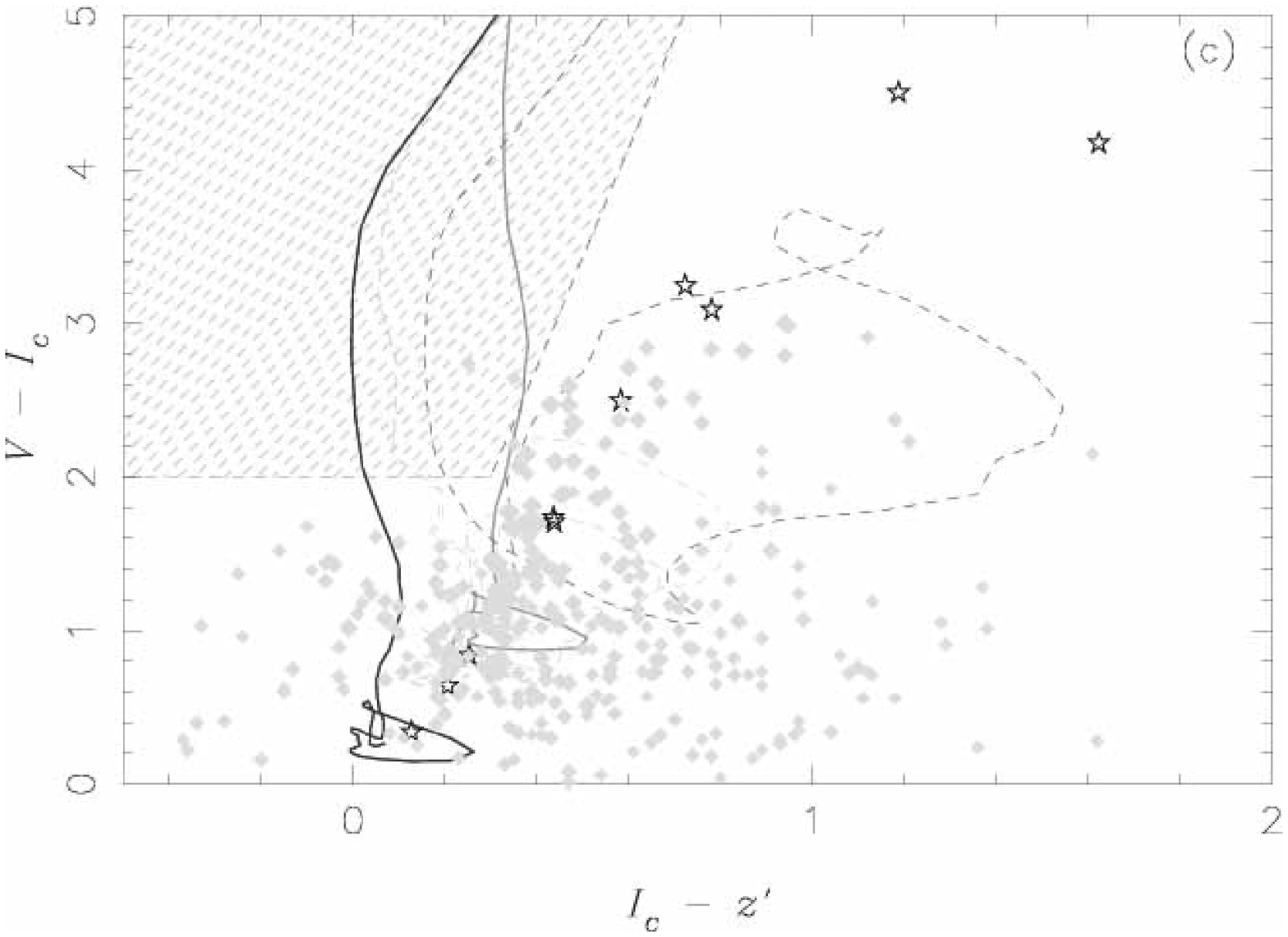}
    \FigureFile(70mm,48mm){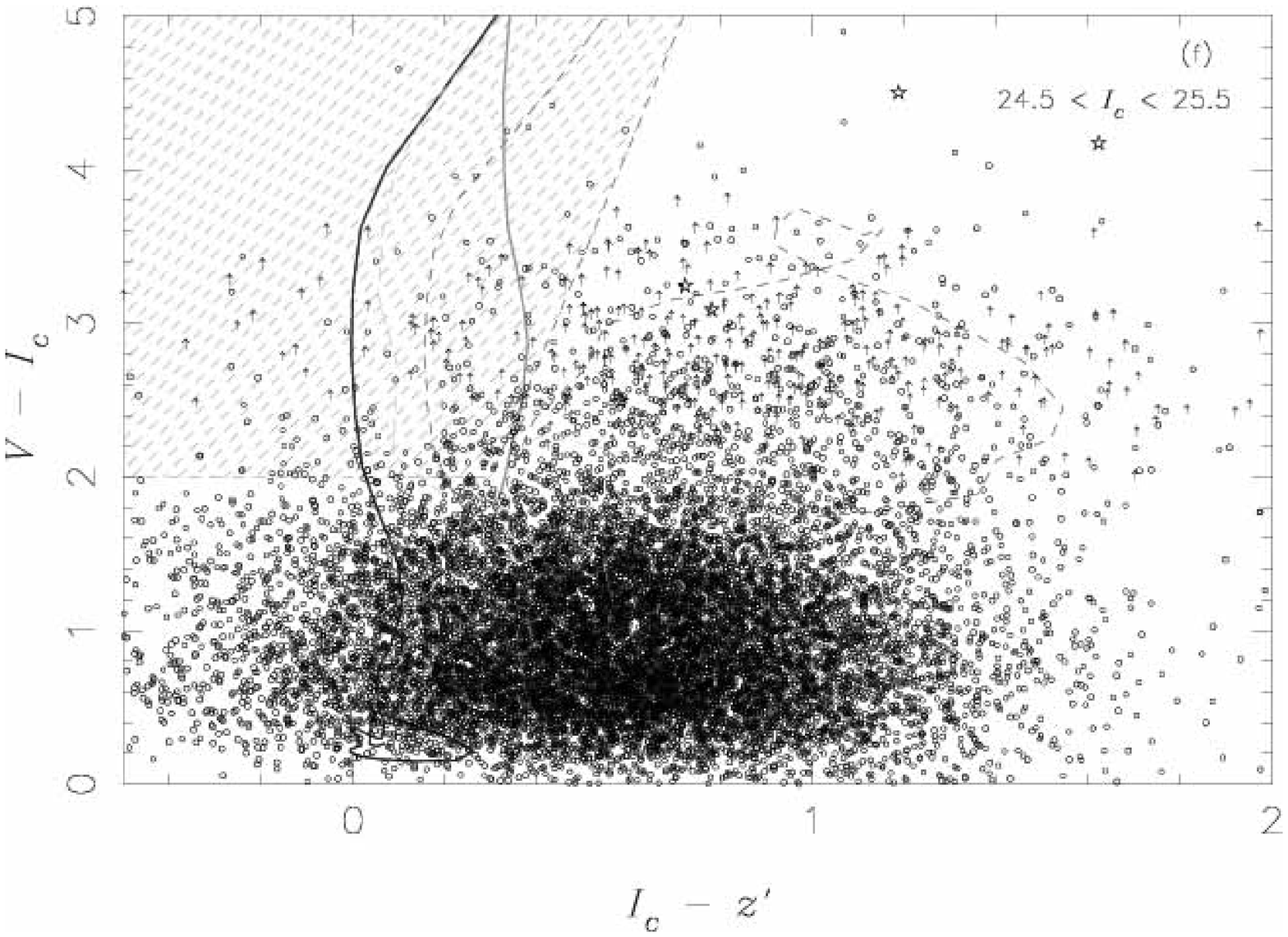}
  \end{center}
  \caption{$I_c - z'$ versus $V - I_c$ diagrams. 
 {\bf (a):} 
 Tracks of colors for model galaxies at redshift ranging 
 from 0 to 5.5.
 A blue (an orange) line represents a galaxy with the bluest (reddest) 
 model spectrum, i.e., $E(B-V)=0.0$ ($E(B-V)=0.4$)
 (see figure \ref{fig_sed}).
 A green and cyan dashed lines refer to 
 local spiral galaxies (green is for Sbc, and cyan is for Scd;
 \cite{Coleman(1980)}). 
 A red dashed line is for a local elliptical galaxy.
 No evolution is considered.
 The symbols are plotted at a redshift interval of 0.5, and the 
 symbols at $z \geq 4.5$ are enlarged.
 Star-symbols indicate the colors of A0 -- M9 stars calculated based on
 the library by \citet{Pickles(1998)}. 
 Hatched region indicates the color criteria 
 we adopted for $z \sim 5$ galaxies.
 {\bf (b):}
 Colors of galaxies cataloged in the spectroscopic
 survey of galaxies in the HDF-N and its flanking fields by
 \citet{Cohen(2000)} and identified in our image.
 All galaxies with the redshift range $z < 4.5$ are plotted.
 Tracks for model spectra are also shown.
 {\bf (c):}
 Same as (b), but for galaxies in the HDF-N for which the photometric
 redshifts are given by \citet{F-Soto(1999)} as $z < 4.5$.
 {\bf (d) -- (f):} 
 The color distribution of the sources extracted from our image. 
 Panels (d), (e) and (f) show objects with $I_c \leq 23.5$, 
 $23.5 < I_c \leq 24.5$ and $24.5 < I_c \leq 25.5$, respectively.
 Circles are objects detected in all of $V$, $I_c$, and $z'$-bands, 
 and arrows are those detected in $I_c$ and $z'$ but 
 not in $V$-band. Lower limits of $V-I_c$ colors are given for
 them. 
}
 \label{fig_twocol}
\end{figure}

A desirable combination of the filters should give the most distinct
separation of the $z \sim 5$ population apart from the objects with
smaller redshifts.
We chose a combination of $V$, $I_c$, and $z^{\prime}$-bands.
In the figure \ref{fig_twocol}a,
we show the two-color $V - I_c$ versus $I_c - z'$ diagram. 
It is seen that galaxies at $z>4.5$ can be well separated
from foreground objects in the diagram.
The most critical contamination is caused by elliptical galaxies
at $z \sim 0.5$ due to the 4000 \AA{} break and early type
spirals at intermediate redshifts.

Selecting $R$-band instead of $V$-band may also be reasonable,
because of the high quantum efficiency at the band.
However, contamination by galaxies with 4000 \AA{} break at
intermediate redshifts is more serious than the adopted filter
combination.
Another reason we did not choose $R$-band is that the continuum 
depletion from the Lyman alpha forest should be heavier in $V$-band, 
and the clear Lyman break makes it easier to isolate LBGs 
in a two-color diagram.

\subsection{Observations}

The observations were carried out on 22nd and 23rd February, 2001 with 
Suprime-Cam
attached to Subaru telescope at Mauna Kea.
The Suprime-Cam consists of 10 mosaiced CCDs, each of which has a
dimension of 2046 by 4090 pixels with a pixel scale of
0.$^{\prime\prime}$2.
Each CCD covers a field of $\sim 7^{\prime}$ by $\sim 13^{\prime}$ and
a typical gap between the CCDs is $10^{\prime\prime}-15^{\prime\prime}$.
The details of the observation are summarized in table 1.
Note that our observations were made before the upgrade of the CCD array
configuration held on April 2001, and the one CCD chip
(at the north-west corner) was not working and another chip
(at the south-west corner) had many bad columns.
Thus 8 CCDs were effectively working, and 
about $\sim 27^{\prime}$ by $\sim 27^{\prime}$ field is 
covered by a exposure.

The imaging observation was made
pointed to the Hubble Deep Field-North (RA(2000) = \timeform{12h36m49s.4}, 
Dec(2000) = +\timeform{62D12'58''}).
Images were taken with a small dithering (typically
20$^{\prime\prime}$) with a typical exposure time of 1440 sec, 
180 -- 240 sec, and 120 sec, 
for $V$, $I_c$, and $z^{\prime}$-bands, respectively for each exposure.
A seeing size during the observations was typically $\sim \timeform{0''.9}$
and we discarded some frames
with a median FWHM of stars larger than $\timeform{1''.0}$ for keeping the
angular resolutions of the final images.
Effective integration times were 21.6 ksec, 2.5ksec, and 3.1 ksec
for $V$, $I_c$, and $z^{\prime}$-bands, respectively.

\begin{table}
\begin{center}
\caption{Details of the observation.}
 \begin{tabular}{lccc}
  \hline\hline
  Dates                  & \multicolumn{3}{c}{22 and 23 Feb. 2001} \\
  CCD dimension          & \multicolumn{3}{c}{9 CCDs with 2046 $\times$ 4090 pixels}\\
  Pixel scale            & \multicolumn{3}{c}{0.2$''$/pixel}\\
  Field of view / Effective surveyed area[$\ast$] & \multicolumn{3}{c}{836 arcmin$^2$ / 575 arcmin$^2$}\\
  Filters                & $V$ & $I_c$ & $z'$ \\
  \cline{2-4}
  Effective integration time & 21,600 sec. & 2,520 sec. & 3,100 sec.\\
  Seeing  & 0$''$.8 -- 1$''$.0 & 0$''$.7 -- 1$''$.0 &0$''$.6 -- 1$''$.0\\
  Magnitude of 5$\sigma$ detection limit[$\dagger$] & 28.5 mag. & 25.5 mag. & 25.2 mag.\\
  \hline
  \multicolumn{4}{@{}l@{}}{\hbox to 0pt{\parbox{120mm}{\footnotesize
  \par\noindent
  \footnotemark[$\ast$] The effective surveyed area represents the area 
  used to derive the luminosity function.
  \par\noindent
  \footnotemark[$\dagger$] The magnitude of the faintest object that 
  can achieve S/N of 5 with the typical aperture size corresponding to 
  the area of the detected sources at the magnitude range 
  (\timeform{0.''9} for $V$ and $z'$-bands, 
  \timeform{1.''2} for $I_c$-band).
  }\hss}}
 \end{tabular}
 \label{tbl1}
\end{center}
\end{table}

\subsection{Data Reduction and Calibration}

The data reduction was carried out in a standard manner using 
IRAF\footnote{IRAF is distributed by the National Optical Astronomy 
Observatories, which are operated by the Association of Universities 
for Research in Astronomy, Inc., under cooperative agreement with the 
National Science Foundation.} for basic data processing;
bias was subtracted using counts in an overscan region, flat field 
frames were made by using normalized average of the object frames.
The sensitivity differences between the CCD chips have been corrected
using the relative count rates of the flat-fielded 
dome flat frames for all three bands.
The background was subtracted by two dimensional polynomial fitting.
Positions of the frames are aligned by measuring the positions of the 
field stars appeared commonly in all frames and fitting fourth-order
polynomial functions for coordinate conversion.
Counts of each frame were adjusted by comparing the count-rate of the 
field stars common to all frames for every filter;
we measured the counts of field stars in each frame, then 
the images were divided by the median of the count ratios 
of stars relative to a fiducial frame so that the count of stars 
are almost constant among frames.\footnote{ 
In this procedure, we found that although the variance of the flux
ratios is stable and is well correlated with the airmass of the
observed sky region on 23rd Feb, the flux ratio varies largely and
the correlation of the flux ratio and the airmass is weaker on 22nd Feb.
This suggests that the sky condition of the second observing night
was better than the first night.
Thus we selected one frame per filter as a fiducial frame, 
which was taken
on the second night and the observed time is close to that of the
exposures for standard stars. 
We adjusted the count-rate of the frames taken on the first night
and other frames taken on the second night according to the
fiducial frame.}
The images were convolved so that the median FWHM of stellar sources 
match within 10\% for all frames in three bands.

Photometric calibrations for $V$ and $I_c$-band images were made using 
images of photometric standard stars in \citet{Landolt(1992)}.
The accuracy was $\sim 0.02$ mag for $V$ and $\sim 0.08$ mag for $I_c$.
For the photometric calibration of the $z$-band images, we employed 
two ways both based on the $I_c - z'$ color.
We first set an arbitrary zero point for $z'$ magnitude. 
In the first method, 
we examined the distribution of $V - I_c$ and $I_c - z'$ colors of field
stars exposed on our images. 
Next we calculated the $I_c - z'$ colors 
to be observed with the system sensitivity of Suprime-Cam
for the stars in the stellar flux library by \citet{Pickles(1998)}.
We can determine the magnitude zero point for $z'$-band by
comparing the distribution of these colors.
Another method is to use the spectral energy distribution of the 
galaxies in the HDF-N and its flanking fields. 
We used the galaxies for which the published $V$, $I$ and $J$-band
magnitudes are available and estimated 
their $I_c - z'$ colors by linear interpolation.
Comparison with  the colors in our image gave the magnitude zero point.
We adopt the magnitude zero point 
determined by the latter method. 
The $z'$ magnitude zero points derived by these two methods agrees 
within $\sim$ 0.1 mag in the color range of $I_c - z'$  between
$\sim 0$ and 1. 
The $z'$-band magnitude is also based on Vega.
We used the flux density table of Vega given by \citet{Castelli(1994)}, 
and applied $z'$(AB) - $z'$(Vega) = 0.5505.

\subsection{Source Detection}

Prior to the extraction of the astronomical objects from the final
imaging data, we set masks for source detection around the bright
objects in order to avoid the misidentification of the noise or
tails of the bright objects as an individual object.
We also removed the edge regions of each CCD chip 
from the object detection because of their low signal-to-noise ratio 
caused by dithering imaging.
In total 26.1 \% of the CCD chip area is masked, and the effective area
observed is 618.6 arcmin$^2$ (575.0 arcmin$^2$ without the CCD chip with
many bad columns). 

We used the software SExtractor \citep{Bertin(1996)} for 
the source detection and photometry.
The primary parameters for the object detection are the threshold of
count above which the pixel is treated as a part of an object and 
the minimum number of connected pixels to be classified as an object.
We extensively examined these parameters by comparing the detected sources 
with the objects in the HDF catalog version 2 \citep{Williams(1996)} 
brighter than $I_{\mathrm{AB}}(814) = 27$ , 
to minimize both the rate of noise contamination and the rate of 
missing objects.
We also tested some spatial smoothing filters and 
adopted a Gaussian smoothing filter with a FWHM of 3 pixels for $I_c$ and
$z'$-bands. 

The best set of parameters with which the detection rate of faint
objects and the noise contamination are compromised 
is that the thresholds are 1.3, 1.0, 1.0 times of
pixel-based background noise 
for $V$, $I_c$ and $z'$, respectively and to have at
least 4 connected pixels above threshold in each band.
We regard sources in different bands as the same object whose pixel 
coordinates are within 1.0$\timeform{''}$.
We pick out objects only if they are detected both in $I_c$ and 
$z'$-bands.
The number of sources thus selected is 29,094 
for $I_c$ $\leq$ 25.0 and 35,362 for $I_c$ $\leq$ 25.5 in
the 9 CCD chips.
Although the adopted detection criteria seem to be loose,
i.e., seem to tend to pick out many noise features,
the criterion of two band detection removes spurious sources efficiently.
It would be clarified by the comparison of the detected objects 
to the HST/WFPC2 data of HDF-N \citep{Williams(1996)}, which have deeper 
$V(606)$ and $I(814)$ images.
There is no misidentification of noise as an object in 
$I_c \leq$ 24.5 and the number of the noise misidentification in 
$24.5 < I_c \leq 25.0$ is two among 55 extracted sources within the 
HDF-N region.
Even in the magnitude range 25 $<$ $I_c$ $\leq$ 25.5, which is the
faintest range we use, the rate of misidentification of noise is not so
large; it is $\sim$ 17\%. Many of these spurious sources locate 
close to the objects which are relatively bright 
(but fainter than masked objects).
These objects degrade quality of flat field around them and 
presumably cause misidentification. 
The number of selected sources are robust for changes of the 
detection parameters; if we adopt 10 pixels for the minimum connection of 
pixels for $V$ and $I_c$-bands, we only lose $\sim 2\%$ 
of the objects at the faintest magnitude range.
The vast majority of selected objects have area larger than 10 pixels 
in $I_c$-bands.
The fraction of selected objects in our image 
among the HDF-N catalog is getting smaller as the magnitude goes fainter;
$\sim$ 100 \% for $I_{\mathrm{AB}}$(814) $\leq$ 25, 
$\sim$ 67 \% for 25 $<$ $I_{\mathrm{AB}}$(814) $\leq$ 25.5,
$\sim$ 28 \% for 25.5 $<$ $I_{\mathrm{AB}}$(814) $\leq$ 26.0.
The detection rates in other CCD chips do not change much as
described in section 4.1.
The number counts in $I_c$-band give 
$4.0\times10^4$ deg$^{-2}$ mag$^{-1}$ for $23.0 < I_c \leq 23.5$, 
$5.6\times10^4$ for $23.5 < I_c \leq 24.0$, 
$7.4\times10^4$ for $24.0 < I_c \leq 24.5$, 
$8.3\times10^4$ for $24.5 < I_c \leq 25.0$ and 
$7.7\times10^4$ for $25.0 < I_c \leq 25.5$.
If we correct the incompleteness of our source detection,
the number counts of our data agree well with those appeared on 
previous studies (e.g., \cite{Tyson(1988)}; \cite{Williams(1996)}).

Photometry of the detected objects was also made using
SExtractor. We use the value ``best magnitude'' in SExtractor 
for $I_c$ magnitude and 
a \timeform{1''.6} diameter aperture magnitude for $V-I_c$ and $I_c-z'$ colors.
The photometric accuracies were estimated by putting many artificial
objects into our final images, executing detection and photometry 
with the same manner for the observed data and compare returned 
value with original magnitude.
The sizes and apparent magnitudes of the artificial objects 
were adjusted so that the obtained magnitude and size distribution 
is similar to that of real data.
The rms errors of $I_c$-band ``best magnitude'' for objects with 
$24.0 < I_c \leq 24.5$, $24.5 < I_c \leq 25.0$
and $25.0 < I_c \leq 25.5$ were 0.28, 0.35 and 0.41 mag, respectively.
These error values are roughly consistent with the errors of $I_c$
magnitude estimated 
from the comparison with the $I_{814}$ magnitude 
for the objects in the HDF-N region.
Photometric errors for $V$-band data are 0.22, 0.31 and 0.33 mag for 
$26 < V \leq 26.5$, $26.5 < V \leq 27.0$ and $27.0 < V \leq 27.5$,
respectively, and for $z'$-band data, 
0.39 mag, 0.45 mag and 0.49 mag for $23.5 < z' \leq 24.0$, 
$24.0 < z' \leq 24.5$ and $24.5 < z' \leq 25.5$, respectively.
The errors in $V-I_c$ and $I_c - z'$ colors, 
which were measured with a \timeform{1''.6} aperture
magnitude, are smaller than the errors in the total magnitude. 
The rms errors of background noise in $I_c$-band, 
which were measured 
at randomly selected positions in our image 
with a \timeform{1''.6} aperture, 
were 0.08 mag and 0.28 mag for $I_c = 24.0$ and $25.5$, 
respectively. For the $V$-band data, they were 0.07 mag and 0.25 mag 
for $V = 26.0$, and $27.5$ respectively, 
and for the $z'$-band data, 
they were 0.18 mag and 0.40 mag for $z' = 24.0$ and $25.0$, 
respectively.

The astrometry for detected objects were made using 
stars in our field-of-view listed in 
APM sky catalog \citep{Irwin(1994)}.
We added stars from USNO-A2.0 \citep{Monet(1998)} to increase the 
number of stars for calculation of conversion coefficients.
There is no systematic error larger than \timeform{0.''3} between 
these two catalogues in our region.
We measured the positions of the stars in our $I_c$-band final image 
and fitted a fourth-order polynomial function to minimize
the $\chi^2$ of the positional errors, for conversion of positions
in our images to equatorial coordinates.
The relative positions of CCD chips were also
adjusted to provide a minimum error for the coordinate conversion.
The achieved internal positional accuracy was $\sim \timeform{0''.4}$
rms for most part of the image, and $\sim \timeform{1''}$ at one
CCD chip with many bad columns at the south-west corner.

\section{Lyman Break Galaxy Candidates at $z \sim 5$}

\subsection{Selection Criteria}

One of the most serious problems in the study of statistical 
properties of Lyman break galaxies is a 
certain amount of contamination by objects located between us and 
objects at the targeted redshift (so called 'interlopers').
The uncertainty of the fraction of the contamination affects the 
reliability of statistical properties of the population,
such as the luminosity function or the amplitude of the clustering.
In figure \ref{fig_twocol}a, we can see that a galaxy with an old
stellar population (red dashed line) at $z \sim 0.5$ is 
close to the selection window for $z \sim 5$ galaxies;
the 4000 \AA\  break in a galaxy at $z \sim 0.5$ mimics the color
by the Lyman break at $z \sim 5$.

In order to determine the selection criteria for galaxies at $z \sim 5$ 
as reliable as possible, 
we have utilized the previously published redshift data of the 
HDF-N and its flanking fields. 
We used the list of galaxies in the HDF-N and the flanking fields 
by Caltech Faint Galaxy Redshift Survey 
(CFGRS; \cite{Cohen(2000)}) and the catalog of photometric 
redshift of galaxies in the HDF-N by \citet{F-Soto(1999)}.
The magnitude limit is $R \sim 25.5$ mag for the CFGRS catalog 
and $I_{\mathrm{AB}}(814) = 28.0$ mag for the photometric redshift catalog.
In our data we detect 566 galaxies out of the 
671 galaxies in the CFGRS catalog and 317 galaxies out of the 477 
galaxies with $I_c < 26$ and at $z < 4.5$ in the photometric 
redshift catalogue.
Most of the galaxies in the CFGRS catalog which have no counterpart in 
our image are located outside of our final image (between CCD chips or 
in the masked regions).
Galaxies listed in the catalog by \citet{F-Soto(1999)} 
but not in our image are objects fainter 
than our limiting magnitude.
In figure \ref{fig_twocol}b and \ref{fig_twocol}c, we plot
the colors ($I_c - z^{\prime}$ and $V - I_c$) of thus identified galaxies 
at $z < 4.5$ in the CFGRS sample and in
the photometric redshift sample, respectively.

Considering the distribution of the foreground galaxies in the
two color diagrams
we adopt the selection criteria for galaxies at $z \sim 5$ as

\begin{eqnarray}
 V - I_c \geq 2.0,
\end{eqnarray}
and
\begin{eqnarray}
 V - I_c \geq 7 (I_c - z') - 0.1.
\end{eqnarray}

\noindent
The selection window defined by these criteria 
is shown as a hatched region in figure \ref{fig_twocol}.
There is no galaxy in the selection window in the spectroscopic
redshift database with $z < 4.5$ 
and only two galaxies in the photometric redshift
database estimated to be at $z < 4.5$.
These two galaxies are HDF 3$-$0238.0 and HDF 4$-$0505.1.
The $I_c$ magnitude of HDF 3$-$0238.0 in our image is 24.8 mag and 
its position in the two color diagram  
is close to the boundary of our selection criteria.
It is classified as an elliptical galaxy at $z = 0.92$
by \citet{F-Soto(1999)} and has a round shape 
both in drizzled HDF-N image and in our $I_c$-band image. 
The estimated redshift of HDF 4$-$0505.1 by \citet{F-Soto(1999)} 
is 1.44. 
It is associated with a more extended object HDF 4$-$0505.0,
for which the photometric redshift is not available
and it has similar colors to HDF 4$-$0505.1.
Our $I_c$-band image cannot resolve these two objects. 
These objects could be interlopers which happen to 
enter our color selection window.

In figure \ref{fig_twocol}d--f, the color distribution of the 
all detected objects is shown.
For objects which are not detected in $V$-band, we assign the
lower limit of $V-I_c$ color using the limiting $V$ magnitude of
28.5 mag.
We regard such objects (detected in $I_c$ and 
$z^{\prime}$-bands, but not in $V$-band) as candidates at $z\sim5$, if they
enter the selection window.
There are objects with $V-I_c$ and $I_c-z'$ colors whose colors do not 
match any type of model galaxies or stars. For example, about 4\% 
of all objects have $I-z'< - 0.1$. The fraction of such objects with 
blue $I-z'$ color increases in fainter magnitude.
Although some of their colors may be affected by emission lines, 
we suppose most of them have redder intrinsic colors but 
are placed due to photometric errors.

\begin{figure}
  \begin{center}
    \FigureFile(150mm,200mm){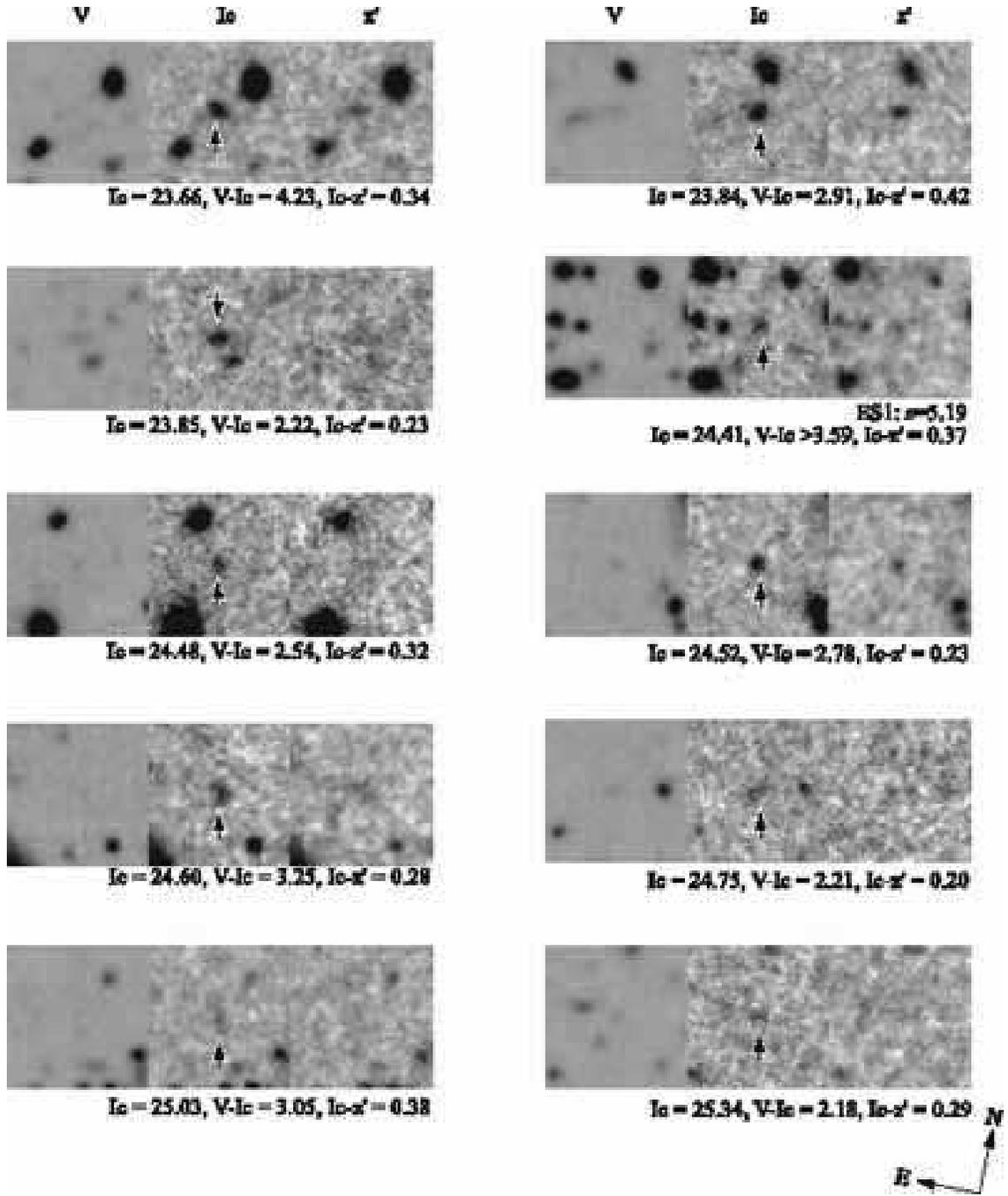}
  \end{center}
  \caption{Images of the representative LBG candidates, in an order 
 of $I_c$ magnitude. From left to right, 
 $V$, $I_c$ and $z'$-band images are shown in each panel. 
 The field of view is \timeform{10''}.
}
 \label{fig_montage}
\end{figure}

Among the 35,362 detected sources, 321 objects with $I_c \leq 25.5$
fall within this selection window.
There are 11 objects with $20.0 < I_c < 23.0$. 
Most of them lie very close to the boundary of
our color selection criteria, and seven objects have 
neighbours within $\timeform{5''}$.
All of them are clearly visible in our $V$-band image and 
have round shapes, suggesting that they are elliptical 
galaxies with old stellar population at intermediate redshift.
They are also unusually luminous if they are star-forming galaxies 
located at $z \sim 5$ (absolute UV magnitude would be $\lesssim -22$).
We regard them as the interlopers which happen to 
fall within our color selection criteria due to photometric 
error and excluded them from our candidate list.
In figure \ref{fig_montage}, we show $V$, $I_c$ and $z'$-band images 
of the representative objects in our LBG candidates 
in an order of $I_c$ magnitude.

\subsection{Cross-identification with Galaxies in the HDF-N and Its Flanking Fields}

\begin{figure}
 \begin{center}
   \FigureFile(150mm,150mm){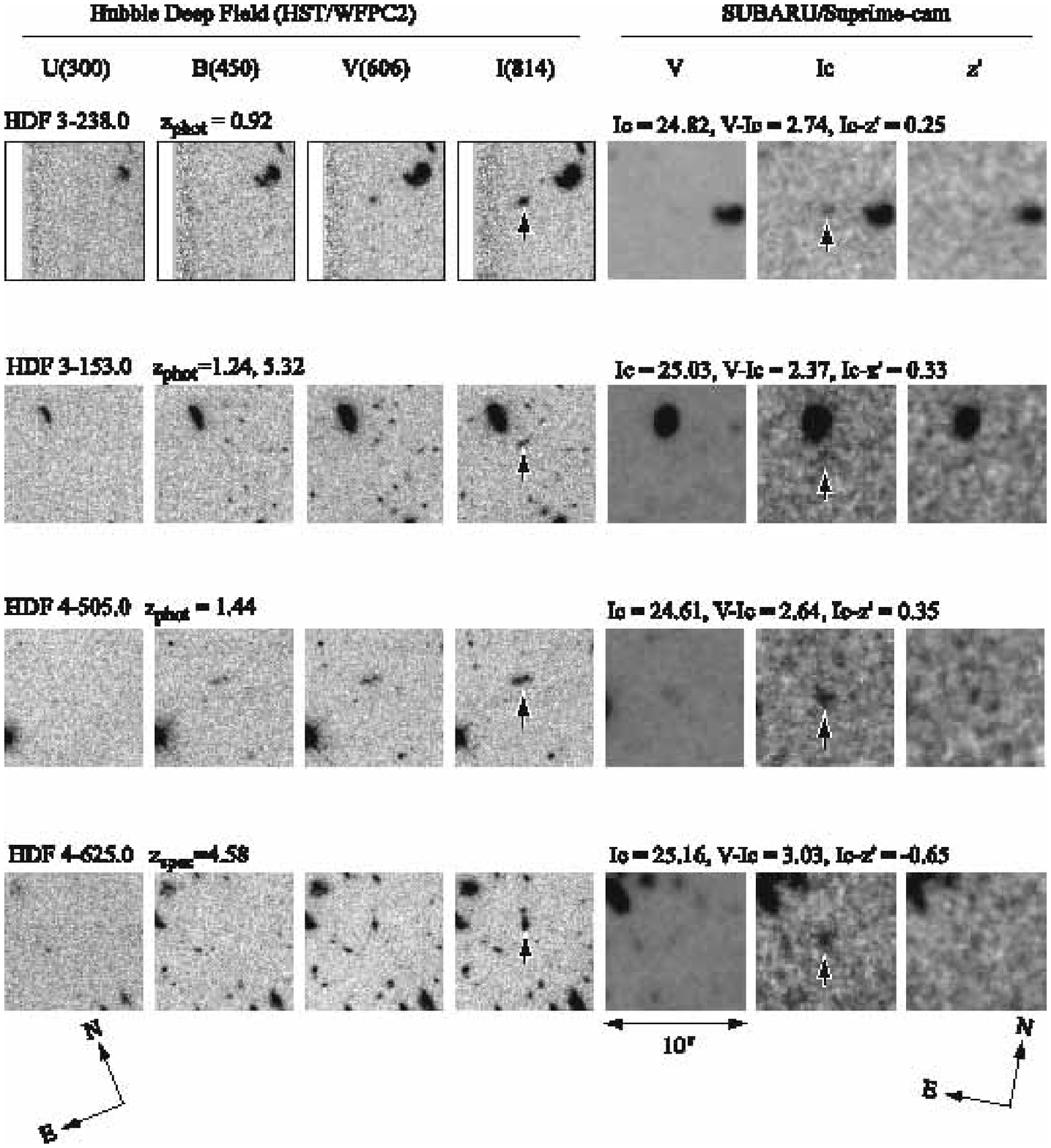}
 \end{center}
 \caption{Images of four LBG candidates in the HDF-N region, of which 
 redshifts (either spectroscopic or photometric) are 
 available. The field of view is \timeform{10''}. Note that position angles
 of HDF/WFPC2 images and Suprime-Cam images are different, as shown in
 the bottom of the figure.
}
 \label{fig_hdfmontage}
\end{figure}

We find six LBG candidates in the HDF-N region.
The range of the $I_c$ magnitude of these objects are from 24.6 to 25.4.
One object HDF 4$-$625.0 is spectroscopically confirmed 
to be at a redshift of 4.58, and 
for three objects photometric redshifts are 
provided by \citet{F-Soto(1999)}.
The appearances of these four objects with redshift information 
in the HDF-N drizzled images, as well as in the Suprime-Cam images, 
are shown in figure \ref{fig_hdfmontage}.
Among the three objects for which photometric redshifts are given, 
two galaxies (HDF 3$-$238.0 and HDF 4$-$505.0) have 
estimated redshifts of 0.92 and 1.44 respectively. 
These objects are detected in 
$B_{450}$-band, so these would be interlopers.
The other object HDF 3$-$153.0, which appears to be a 
single object in our ground-based image, is identified 
as two resolved objects by \citet{F-Soto(1999)}; 
one object has a redshift of 1.24 and the other 5.32.
Since the latter object is 0.9 mag brighter than the former 
in $I_{814}$-band, the detection by our selection criteria 
is presumably attributed to the object at higher redshift.
As we will describe in the next section, the fraction of 
interlopers at this magnitude range is expected to be 
around 40 to 50 \%. Thus it is not surprising that half of 
the objects with redshift information seems to be interlopers.
The remaining two objects lie close to the edges of the HDF-N 
and photometric redshift measurement is not performed on them.

In the flanking fields of the HDF-N we find 12 LBG candidates.
Two galaxies have been previously identified spectroscopically 
as objects at $z \sim 5$.
They are J123649.2+621539 (ES1),
a serendipitously detected object at $z$ = 5.19 \citep{Dawson(2002)}, 
and B01$-$174, a broad-line AGN at $z = 5.186$ discovered in 
the Chandra Deep Field North region by \citet{Barger(2002)}.
The images of ES1 are shown in figure \ref{fig_montage}.

Vice versa, there are 7 galaxies in the HDF-N and its flanking fields, 
which have been spectroscopically identified as $4.4 < z < 5.5$ but 
are not included in our candidate list.\footnote{There are three objects
with a spectroscopic redshift larger than 5.5 in the HDF-N and its flanking
fields. All of them are faint and 
not detected in our image.  As it is described in the next section, 
incompleteness of our data gets larger for objects at $z \gtrsim$ 5.5.}
One object is located close to the edge of a CCD chip and the source
detection was not executed there. Five galaxies are fainter 
than our magnitude limit or spatially unresolved with a 
nearby object in our image.
The remaining one object is HDF 4$-$439.0, of which redshift is 
given as 4.54 by \citet{Stern(1999)}. 
The $V-I_c$ and $I_c-z'$ colors of the object are close to the boundary 
of our selection criteria, and we would fail to select the object 
probably due to the photometric error.

The detection rate of our survey for the objects with 
$I_{\mathrm{AB}}$(814) $\gtrsim$ 25.0 with spectroscopic 
redshift between $4.4 < z < 5.5$  
in the HDF-N and its flanking field is $\sim$1/3.
It is consistent with the incompleteness estimated by the Monte Carlo
simulations described in the next section.

\subsection{Morphology and Size Distribution}

In regard to the effect of the merging event on the star formation
activity of the LBGs, 
it is interesting that the HDF-N $I_{814}$ image of HDF 4$-$625.0 
at $z=4.58$ shows the sign of merging or interaction 
(see the bottom panels in figure \ref{fig_hdfmontage}).
If we can figure out the rate of mergers in our sample, 
we might be able to assess the effect of merging events on the 
star formation of the LBGs.
Although our ground-based images have poor angular resolution 
and is not as deep as HDF-N HST/WFPC2 images, 
we try to examine the fraction of galaxies showing 
a sign of interaction with other galaxies among 
our LBG candidates. We count the number of objects 
which have multiple cores in $I_c$ and $z'$-bands 
and are not seen or significantly faint in $V$-band 
among our LBG candidates with $I_c$ magnitude between 
24 and 25 (An example is the object with $I_c=24.60$ in figure 
\ref{fig_montage}).
Roughly 10 \% of 
the LBG candidates show such feature.
This fraction should, however, be regarded as a lower limit,
since we cannot distinguish close encounters and mergers from a 
single object.
The seeing condition of our observation ($\sim 1''$, corresponding to 
$\sim$ 4.4 $h^{-1}$ kpc for $z=5.0$, $\Omega_\mathrm{M}=0.3$ and
$\Omega_\Lambda=0.7$) is not suitable for
the detailed analysis of the sizes and morphologies of the LBG
candidates; about 40 \% of the candidates have half light radius
comparable to the seeing size.

\subsection{Magnitude and Color Distribution}

\begin{figure}
  \begin{center}
    \FigureFile(80mm,50mm){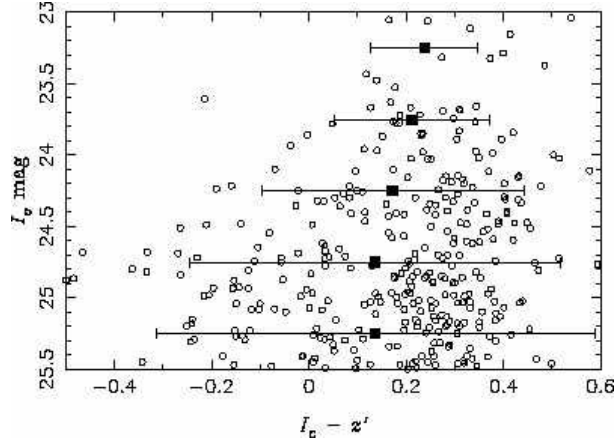}
  \end{center}
  \caption{Apparent $I_c$ magnitude against $I_c - z'$ color 
 for LBG candidates at $z \sim 5$. 
 Mean values of $I_c-z'$ colors in 0.5 $I_c$ mag step are shown 
 with filled squares.
 Each error bar represents the photometric error of the $I_c - z'$ color 
 in each 0.5 magnitude bin, measured by a simulation using artificial 
 objects (see section 2.4).
 
}
 \label{fig_col_mag}
\end{figure}

\begin{figure}
  \begin{center}
    \FigureFile(80mm,50mm){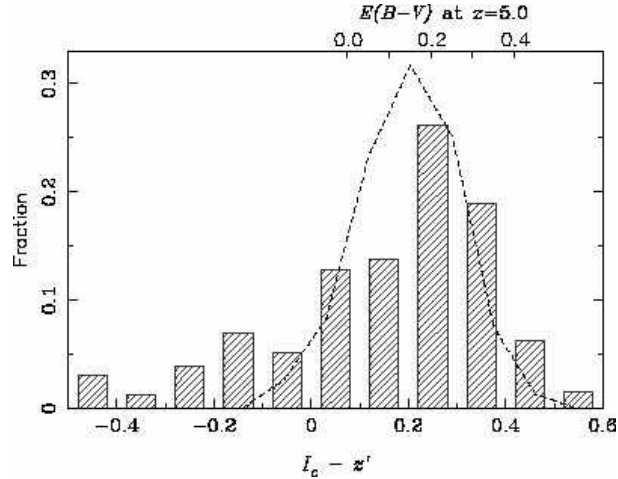}
  \end{center}
  \caption{$I_c - z'$ color distribution of LBG candidates at $z \sim 5$.
 The objects with $I_c - z'$
 colors less than $-0.5$ are included in the bin for the bluest color.
 The ticks at the upper
 horizontal axis indicate the expected $I_c - z'$ colors of a
 model galaxy at $z = 5$ with $E(B-V)$ from  0.0 to 0.4.
 See text for the details. 
 The dashed line shows a distribution of $E(B-V)$ for  LBGs
 at $z \sim 3$ by \citet{S99}.
 }
  \label{fig_color}
\end{figure}

In figure \ref{fig_col_mag}, 
apparent $I_c$ magnitudes against the color ($I_c - z^{\prime}$) 
are plotted.
A tendency that bright blue candidates are deficient is seen.
Other any clear trends are not seen, though 
the photometric error is larger for faint objects
($\sim$0.45 mag in $25 < I_c < 25.5$).
Figure \ref{fig_color} shows a distribution of $I - z'$ colors of
the LBG candidates with $I_c \leq 25.0$ in our search.
The $I_c - z'$ color of 0.07 mag corresponds to the
bluest model LBG spectrum we adopted in section 2.1,
which is regarded here as the extinction free template LBG spectrum. 
If the redder $I_c - z'$ colors originate in the internal
extinction, we are able to derive a rough estimate of the extinction
of each galaxy by comparing with the model spectra with different
degree of extinction. 
In the upper abscissa of the figure, we indicate the $I_c - z'$ color
of the model spectra of LBGs at $z = 5$ with $E(B-V) = 0.0$ to 0.4
assuming the extinction law by \citet{Calzetti(2000)}.
As it is seen in figure \ref{fig_twocol}a, the relationship between 
$I_c-z'$ color and dust extinction depends on redshift of an object.
If we adopt redshift 4.8 instead of 5.0, $E(B-V)=0.0$ corresponds to 
$I_c-z' \sim 0$.
The distribution of $E(B-V)$ of the LBGs at $z \sim 3$ estimated by
\citet{S99} is also shown in figure \ref{fig_color} (dashed curve).
The distribution of $E(B-V)$ based on the template fitting for
rest-frame UV to optical SEDs of LBGs at $z \sim 3$
\citep{Papovich(2001)} is broadly consistent with that of \citet{S99}.
Although the photometric errors for fainter objects in our candidates
are large 
and a possible contamination by the
emission lines might disturb this distribution, the degree of the 
dust extinction for LBG candidates detected by our search
seems to be similar to those at $z \sim 3$.

\section{The Luminosity Function of the LBG candidates at $z$ $\sim$ 5}

In this section, we try to derive a luminosity function of LBGs at
$z \sim 5$ {\it statistically} using our LBG candidates.
The redshifts of the candidates are not yet confirmed by
spectroscopy except for a few objects.
However, considering that the method works so efficiently
at $z \sim 3$ and $z \sim 4$ and the color selection criteria
adopted here is well examined thanks to a large redshift database
based on deep spectroscopic and multi-waveband photometric observations
in HDF-N and its flanking fields,
we believe the catalog is reliable enough to
discuss statistical nature of them.
We would like to make follow-up spectroscopy in future,
but it would take long years.
Under these circumstances, constructing luminosity function and
discussing its implication to cosmic star formation
history statistically is opportune at this stage.
In deriving the luminosity function of the LBGs at $z \sim 5$,
we should consider the two important issues; 
contamination by interlopers and incompleteness of the sample.
In the following procedure of this section 
we exclude the data in the CCD chip at the south-west
corner with many bad columns; only five candidates are detected
in the chip. 

\subsection{Correction for Contamination by Interlopers} 

In section 3 it has been described that our selection criteria 
are defined to be least contaminated by interlopers via redshift 
database of galaxies in the HDF-N and its flanking fields.
While the galaxies in the HDF-N and its flanking fields are quite 
useful for the determination of the selection criteria, 
they are not suitable for the estimation of a number of 
interlopers contained in our LBG candidates, 
because the number of these galaxies with available redshift 
is only a few per cent of all the detected sources in 
our survey. 
Thus the estimation of the contamination rate based on 
the whole sample of detected objects in our catalog 
is necessary for the proper correction 
for interlopers.

There are some possible causes of contamination by interlopers 
other than the photometric errors:
the first one is interlopers with colors which are intrinsically similar
to LBGs at $z > 4.5$, and the second one is systematic errors due to 
the indirect $z'$-band calibration (section 2.3).
The fact that there is no object at intermediate redshift 
for galaxies in the CFGRS and only two among 317 galaxies 
in the photometric redshift catalog of HDF-N by \citet{F-Soto(1999)} 
(see section 3.1 and figures \ref{fig_twocol}b and \ref{fig_twocol}c) 
would certify the validity of our selection criteria (equation (1) 
and (2) in section 3.1). 
This estimation of contamination by interlopers described 
below is based on this assumption.
In regards to the systematic error due to $z'$-band calibration, 
it does not affect to our results, 
because we have defined our color selection criteria 
based on the colors of galaxies at the intermediate redshift in the 
HDF-N and its flanking fields. 
If $I_c-z'$ color distribution is shifted due to the 
systematic zero point shift in $z'$-band, 
the color selection criteria should also be shifted in the same direction.
So we only consider photometric 
errors in quantifying the degree of contamination below.

The  fraction of interlopers varies with 
their apparent magnitude, because the larger photometric errors in
the measurement of fainter objects should increase the probability of 
their invasion into the $z \sim 5$ selection window even if their
intrinsic color is the same as the brighter ones.
Thus we should estimate and correct the fraction of the interlopers
at each magnitude bin.
In order to accomplish this task, we employed
a resampling (bootstrap) algorithm.
First, we selected objects at the outside of the selection window
in the $V - I_c$ and $I_c - z'$ color diagram from all of the 
detected objects.
Then we divided them by their $I_c$ magnitude with a 0.5 mag
step.
For each magnitude bin, we resampled objects randomly 
allowing the possibility of duplication. The number of objects 
included in a resampled group was the same as the number of objects in the
magnitude bin.
Then we assigned photometric errors to $V-I_c$ and $I_c-z'$ colors 
according to the observed $I_c$ magnitude and color of each object; 
thus also according to $V$ and $z'$-band magnitude.
The error values in each band were based on the estimation 
by putting many artificial objects, and 
the obtained error values are 
consistent with those estimated from the background noise 
with \timeform{1''.6} diamter (see section 2.4).
The typical values for objects with colors close to the 
border of our color selection criteria were 
0.07 mag and 0.11 mag for $V-I_c$ and $I_c-z'$ 
colors, respectively, for objects with $23.0 < I_c \leq 23.5$, and 
0.43 mag and 0.45 mag for $V-I_c$ and $I_c-z'$
colors, respectively, for objects with $25.0 < I_c \leq 25.5$.
The error assignment was made randomly under the assumption that 
photometric errors have a Gaussian distribution.
We counted the number of objects of which modified colors are
in the selection window.
We ran this procedure 1,000 times for each magnitude bin and 
we regard the average of the numbers of objects that 
fall in the selection window as the estimated number of interlopers.
We also performed the same procedure 
by adding objects which are placed inside of the selection window 
but are close to the boundary; their color differences 
from the boundary of the selection criteria are within the photometric
errors in both $V-I_c$ and $I_c-z'$.
By this test we intend to estimate a possible maximum number of 
interlopers.
The number of objects within the window after adding the 
random error was roughly twice the number resulted in  
the first test (in which the objects at the outside of the 
selection window are used) over the all magnitude ranges.
Since some (not all, but unknown) fraction of the objects within the
selection window is considered to be interlopers, 
the first test should underestimate the number of interlopers, while 
the second test should overestimate it.
We took the average of the results of these two tests as 
the expected number of interlopers.
The variation of the number densities caused by this uncertainty 
in the estimation of the interlopers is $\sim 10$ -- 30 \% 
for all magnitude bins.
We show the number of the interlopers thus obtained 
in table \ref{tbl_numbers}.
The estimated fraction of interlopers gradually increases 
along with the magnitude,  from 22 \% at 
$23.0 < I_c <23.5$ to 48 \% for $25.0 < I_c < 25.5$.
The result is consistent with the fact that no or only a few
interlopers are seen for spectroscopic or photometric
redshift catalog obtained in the HDF-N and its flanking fields 
(figure \ref{fig_twocol}b 
and \ref{fig_twocol}c), since the area of the HDF-N is smaller than 
our effective surveyed area by a factor of $\sim$ 100.

Subtraction of these numbers from the numbers of objects within the
color selection window gives the expected numbers of objects
at $z \sim 5$ if there is no contamination of noise.
As described in section 2.4, the cross-identification of the
sources with the HDF-N catalog shows that in the magnitude range
$I_c \leq 24.5$ the sources are almost noise-free.
In the fainter magnitude range, we have to multiply the expected
number (difference of the values in columns 2 and 3 in 
table \ref{tbl_numbers})
by the fraction of the real objects among the total list of detected
sources (1 minus the values in the fourth column 
in table \ref{tbl_numbers}), because the fraction of
noises for the detected sources outside of the color selection window
must be the same as that in the window.
Accordingly, we obtain $N(m)$, the corrected number of objects 
at $z \sim 5$ in a magnitude range 
between $m \pm 0.25$.

\begin{table}
\begin{center}
\caption{The number of objects detected in our survey with colors 
 expected for galaxies at $z \sim 5$, and corrections
 for interlopers, noise misidentification, and incompleteness.}
 \begin{tabular}{lrrlrrr}
  \hline\hline
  Magnitude     & Detected & Expected Number & Noise & \\
  range ($I_c$) & Number   & of Interlopers  & Rate[$*$] & \multicolumn{3}{c}{$N(m)$}\\
  (1)           & (2)      & (3)             & (4)   & \multicolumn{3}{c}{((2)$-$(3))*(1$-$(4))}\\
  \hline
  23.0 -- 23.5 &  6&  1.3 (22\%)  & 0.0  & &  4.7 & \\
  23.5 -- 24.0 & 25&  6.6 (26\%)  & 0.0  & & 18.4 & \\
  24.0 -- 24.5 & 61& 18.5 (30.3\%)& 0.0  & & 42.5 & \\
  24.5 -- 25.0 & 91& 33.9 (37.3\%)& 0.036& & 55.0 & \\
  25.0 -- 25.5 &122& 58.5 (48.0\%)& 0.172& & 52.6 & \\
  \hline
  \multicolumn{4}{@{}l@{}}{\hbox to 0pt{\parbox{85mm}{\footnotesize
  \par\noindent
  \footnotemark[$*$] Fraction of misidentification of noise as an
  object, estimated by the cross-identification with the HDF-N catalog.
  }\hss}}
 \end{tabular}
 \label{tbl_numbers}
\end{center}
\end{table}

\subsection{Correction for Incompleteness}

The other thing that we have to consider for the proper derivation
of the luminosity function is the correction for the incompleteness of 
our survey.
We should consider two aspects of incompleteness
\citep{S99}.
One is the detection rate against the magnitude;
the detection rate decreases with increasing magnitude, which
is the usual detection incompleteness.
The other one is the rate against the redshift and color;
as seen in figure \ref{fig_twocol}a, 
at a redshift range around 4 -- 4.5, the detection rate 
should strongly depend on the color of an object.
In addition, at the higher redshift (say $z > 5.5$),
the location of the detected sources in figure \ref{fig_twocol} 
goes out of the window when they are fainter than the detection 
limit in $V$-band.

In order to correct these incompleteness, we ran Monte Carlo
simulations.
These Monte Carlo simulations enable us to calculate the ``effective
volume'' of our survey \citep{S99},

\begin{equation}
V_{\mathrm{eff}}(m) = \int dz {} p(m,z) dV/dz,
\end{equation}

\noindent
where $p(m,z)$ is a probability of finding an object with a magnitude of
$m$ at a redshift of $z$, which can be obtained as the average weighted 
over the distribution of SEDs of LBGs.

\begin{figure}
  \begin{center}
    \FigureFile(80mm,50mm){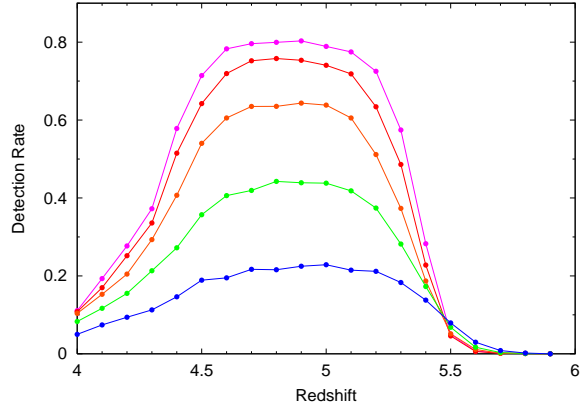}
  \end{center}
 \caption{The detection rate $p(m,z)$ of the galaxies against 
 the redshift of the objects estimated based on the Monte
 Carlo simulations.
 Each line indicates the detection rate for an $I_c$ magnitude
 range of 23.0 -- 23.5, 23.5 -- 24.0, 24.0 -- 24.5, 
 24.5 -- 25.0 and 25.0 -- 25.5 (from top to bottom).
 }
  \label{fig_sel}
\end{figure}

First we created sets of artificial objects with various 
magnitudes and various colors at various redshifts. 
The number of objects created in each set is 200. 
The magnitude range was taken from 23 mag to 25.5 mag with a 0.5
magnitude bin.
Their angular sizes were adjusted so that the obtained 
size distribution of the
model objects resembles that of real objects at the same magnitude.
Their $V - I_c$ and $I_c - z'$ colors of LBGs at $z = 4.0 - 5.9$
were calculated using the model spectra reproducing those of
LBGs
described in section 2.1.
The redshift step was 0.1.
We took objects with five different colors, which correspond
to $E(B-V)$ of 0.0, 0.1, 0.2, 0.3, and 0.4 for each redshift 
and magnitude using the model SEDs.
Thus we created 5 $\times$ 20 $\times$ 5 = 500 sets of 
200 artificial objects for each of 8 CCDs.
Then we put these artificial objects into our $V$, $I_c$ and $z'$-band images
with a random spatial distribution, and performed the 
object detection in the same way as we did for real objects.
Next we counted the number of detected objects which fell in 
our selection window for $z \sim 5$ objects. 
The fraction of the objects which satisfied our color selection criteria 
can be regarded as the probability of detecting an object with
the magnitude and the color at the redshift.
Weighted average of the probabilities are calculated using the 
$E(B-V)$ distribution of LBGs at $z \sim 3$ by \citet{S99} within the
range of $E(B-V) = 0.0$ -- 0.4 (see dashed curve of Fig. \ref{fig_color}). 
As it is described in section 3.4, $E(B-V)$ distribution of our 
LBG candidates is suggested to be similar to that by \citet{S99}.
The resulting probability $p(m,z)$  averaged over 8 CCD
chips is shown in figure \ref{fig_sel}.\footnote{The variation
of $p(m,z)$ among 8 CCD chips is not large; the variation of the 
$V_{\mathrm{eff}}$ obtained by equation (3) for each CCD is 
a few \% for the bright ($23.0 < I_c \leq 24.0$) sources, and 
$\sim$ 35 \% for the faintest sources.
}
The overall decrease in detection rate along the magnitude is clearly
seen.
At the lower redshift ($z = 4 - 4.5$), the detection rate
decreases because of the cutoff of the selection window;
the degree of the decline depends on the color and thus the
color distribution of LBGs.
At the higher redshift ($z=5.2-5.6$), the detection rate
decreases because the sources detected in $I_c$-band but not in
$V$-band go out of the selection window at the fainter $I_c$
magnitude and at the redder $I_c - z^{\prime}$ color.
The sum of the effective volumes of 8 CCD chips obtained for
each magnitude range is summarized in table \ref{tbl_density}
for the two cosmological models.

\begin{table}
\begin{center}
\caption{Number densities of LBG candidates at $z \sim 5$.}
 \begin{tabular}{lccccc}
  \hline\hline
  Magnitude & &\multicolumn{3}{c}{$\Omega_{\mathrm{M}}$ = 0.3, $\Omega_\Lambda$ = 0.7}\\
  \cline{3-5}
  range ($I_c$) & $n(m)$[$*$] & $M_{\mathrm{UV}}$(AB)[$\dagger$] $-$ 5 log $h$ & $V_{\mathrm{eff}}$[$\ddagger$] & log $\Phi(m)$[$\S$]\\
  \hline
  23.0 -- 23.5 & 0.82$\pm$ 0.41 & $-$22.1 -- $-$21.6 & 741.32 & $-4.65$  \\
  23.5 -- 24.0 & 3.20$\pm$ 0.93 & $-$21.6 -- $-$21.1 & 679.07 & $-4.03$  \\
  24.0 -- 24.5 & 7.39$\pm$ 1.83 & $-$21.1 -- $-$20.6 & 565.47 & $-3.58$  \\
  24.5 -- 25.0 & 9.57$\pm$ 2.10 & $-$20.6 -- $-$20.1 & 388.65 & $-3.31$  \\
  25.0 -- 25.5 & 9.14$\pm$ 2.79 & $-$20.1 -- $-$19.6 & 204.31 & $-3.05$  \\
  \hline
\\
  \hline\hline
  Magnitude & &\multicolumn{3}{c}{$\Omega_{\mathrm{M}}$ = 1.0, $\Omega_\Lambda$ = 0.0}\\
  \cline{3-5} 
  range ($I_c$) & $n(m)$[$*$] & $M_{\mathrm{UV}}$(AB)[$\dagger$] $-$ 5 log $h$ & $V_{\mathrm{eff}}$[$\ddagger$] & log $\Phi(m)$[$\S$] \\
  \hline
  23.0 -- 23.5 & 0.82$\pm$ 0.41 & $-$21.2 -- $-$20.7 & 174.01 & $-4.02$ \\
  23.5 -- 24.0 & 3.20$\pm$ 0.93 & $-$20.7 -- $-$20.2 & 159.40 & $-3.40$ \\
  24.0 -- 24.5 & 7.39$\pm$ 1.83 & $-$20.2 -- $-$19.7 & 132.74 & $-2.95$ \\
  24.5 -- 25.0 & 9.57$\pm$ 2.10 & $-$19.7 -- $-$19.2 &  91.21 & $-2.68$ \\
  25.0 -- 25.5 & 9.14$\pm$ 2.79 & $-$19.2 -- $-$18.7 &  47.94 & $-2.42$ \\
  \hline
  \multicolumn{5}{@{}l@{}}{\hbox to 0pt{\parbox{150mm}{\footnotesize
  \par\noindent
  \footnotemark[$*$] Surface number density of LBG candidates in 
  10$^{-2}$ arcmin$^{-2}$ per 0.5 mag. 
  These are derived by dividing $N(m)$ in table
  \ref{tbl_numbers} by the effective surveyed area (575.0 arcmin$^2$).
  \par\noindent
  \footnotemark[$\dagger$] Absolute magnitude in ABmag ($I_c$(AB) =
  $I_c$(Vega) + 0.45), calculated from $I_c$ magnitude by 
  assuming the redshift of 5.0 for all the objects.
  \par\noindent
  \footnotemark[$\ddagger$] Effective volume of our survey in the unit of
  h$^{-3}$ Mpc$^3$ arcmin$^{-2}$. 
  \par\noindent
  \footnotemark[$\S$] Number density of the $z \sim 5$ LBG
  candidates in the unit of h$^3$ mag$^{-1}$ Mpc$^{-3}$.
  }\hss}}
 \end{tabular}
 \label{tbl_density}
\end{center}
\end{table}

\subsection{Luminosity Function at $z \sim 5$} 

\begin{figure}
  \begin{center}
    \FigureFile(80mm,50mm){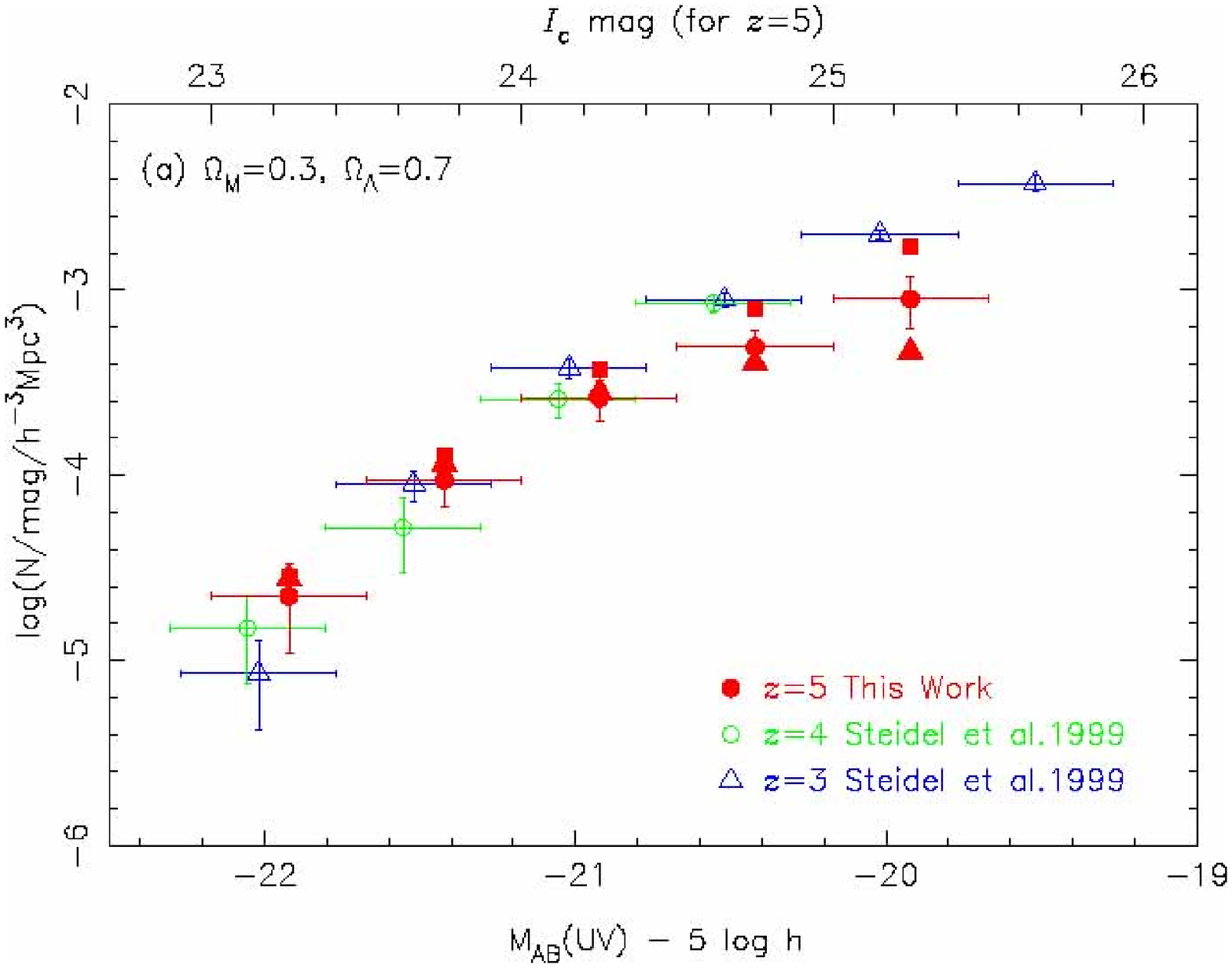}
    \FigureFile(80mm,50mm){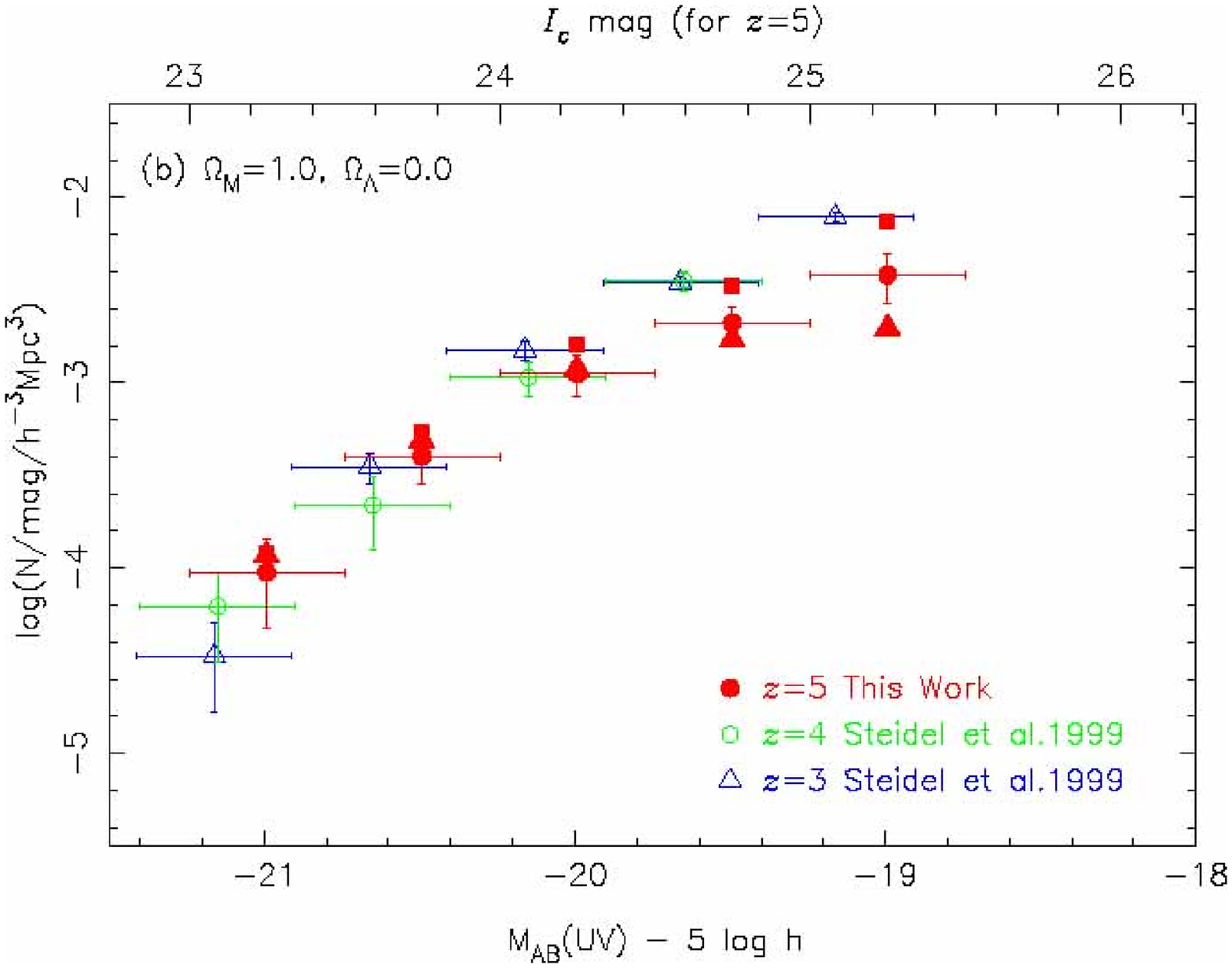}
  \end{center}
 \caption{The luminosity function of the Lyman break galaxies at $z \sim 5$ (red).
 Assumed cosmology is $\Omega_\mathrm{M} = 0.3$ 
 and $\Omega_\Lambda =0.7$ for figure \ref{fig_lf1}a and 
 $\Omega_\mathrm{M} = 1.0$ and 
 $\Omega_\Lambda =0.0$ for figure \ref{fig_lf1}b. 
 The vertical error bars for our $z \sim 5$ data include errors in
 corrections for interlopers and incompleteness, as well as the
 statistical errors in the number of the candidates assuming the Poisson
 distribution.  The assumed redshift is $z = 5.0$.
 The horizontal error bars indicate the magnitude step (0.5 mag) for all
 data points.
 The solid squares show the luminosity function without the correction 
 of contamination, and the solid triangles are for a simple 
 selection function which is 1.0 for a redshift range 4.4 to 5.2 and 
 otherwise 0 regardless of magnitude.
 The open triangles and open circles represent the
 luminosity functions of LBGs at $\langle z \rangle = 3.04$ and 
 $\langle z \rangle = 4.13$, respectively,
 taken from \citet{S99}.
 The vertical error bars for them only show statistical errors.
 }
  \label{fig_lf1}
\end{figure}

The information on the contamination of the interlopers and the
incompleteness of our search described in the previous two subsections 
allow us to calculate the luminosity
function of the Lyman break galaxies at $z \sim 5$ as,
\begin{equation}
\Phi(m) = N(m) / V_{\mathrm{eff}}(m),
\end{equation}

\noindent
where $N(m)$ is the number of objects in a magnitude bin of $m$ 
corrected for the estimated contamination by
interlopers and noise.
The luminosity function thus derived is that at rest UV 1340 \AA,
which redshifts into the central wavelength of $I_c$-band
(8060 \AA) for a redshift of 5.
In order to compare it with the luminosity functions at $z\sim3$ and
$z\sim4$ obtained by \citet{S99}, 
we calculate
absolute AB magnitudes at 1700 \AA\ and derive the luminosity function
at the wavelength, assuming a flat spectrum (in $f_{\nu}$) which
corresponds to $I_c - z'$ color of 0.1 mag in Vega magnitude system.
Table \ref{tbl_density} shows the number density of the LBG candidates
in the 0.5 mag step corrected for the incompleteness and the contamination. 
The luminosity function of the $z \sim$ 5 population is presented
as red circles in figure \ref{fig_lf1}.
In deriving the luminosity function, we assume the redshift of
5.0 for all the LBG candidates to calculate their luminosity distance.
If we use the weighted mean redshift of 4.8 derived from $p(m, z)$ 
(see figure \ref{fig_sel}) instead, 
the absolute magnitude becomes 0.10 mag fainter than  that at $z = 5.0$.
In order to show how much the corrections for contaminations 
and incompleteness affect the luminosity function, 
we show two alternative number densities with solid squares 
and triangles in figure \ref{fig_lf1}. 
Solid squares represent the number density of LBGs at $z \sim 5$ 
in the case that the correction of contamination is skipped; 
i.e., all objects match with 
our color selection criteria are assumed to be genuinely at $z \sim 5$.
Solid triangles indicate the number density 
assuming that the detection rate is 1.0 within a 
redshift range between 4.4 and 5.2 and otherwise 0 
for all magnitude range instead of using the detection rates estimated 
by Monte Carlo simulation shown in figure \ref{fig_sel}.
In figure \ref{fig_lf1} we also show the luminosity functions 
for $z \sim 3$ and 4 obtained by \citet{S99} as open circles 
and open triangles. 
In deriving these luminosity functions, \citet{S99} assume 
$\langle z \rangle = 3.04$ and $\langle z \rangle = 4.13$ 
respectively, according to the 
results of their spectroscopic observations.

There is no significant difference between the luminosity functions for 
$z \sim 3$, 4 and 5 in the bright part.
Although the change of adopted cosmology makes a slight difference
in the relative position of the data points, the general tendency
remains in both figures.
In the fainter part, 
there is a hint of the decrease of the number density at $z \sim 5$.
It should be worth noting here that in figure \ref{fig_lf1}, 
while the vertical error bars for our $z \sim 5$ data include errors in
corrections for interlopers and incompleteness as well as the
statistical errors in the number of the candidates assuming the Poisson
distribution, the error bars for $z \sim 3$ and 4 by \citet{S99} 
only include the statistical errors.
Since the incompleteness of the $z \sim 3$ survey is reported to be 
comparable to or slightly better than ours \citep{S99}, 
if the uncertainty in the corrections for 
incompleteness and contamination of the interlopers 
are considered, the error bars for $z \sim 3$ data points 
should be larger than those appeared in figure \ref{fig_lf1}.
Thus within the current depth and uncertainties of the observed data
it is hard to conclude whether the discrepancy between 
the number density for $z \sim 5$ and that for $z \sim 3$ 
at fainter part is real or not.

\begin{figure}
  \begin{center}
    \FigureFile(80mm,50mm){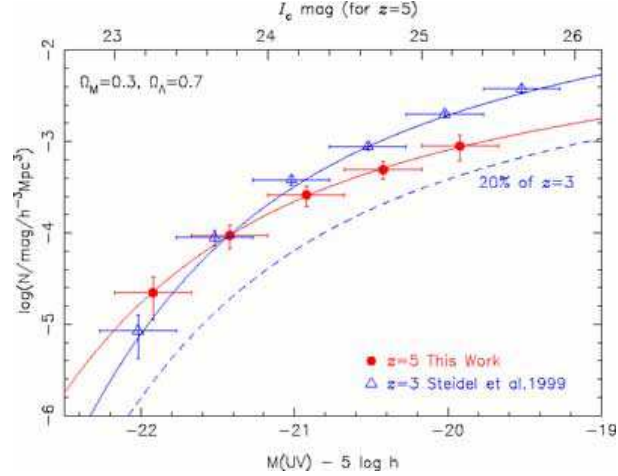}
  \end{center}
 \caption{Same as figure \ref{fig_lf1}a but only for LBGs at $z \sim 3$ 
 and 5. Schechter functions fitted to the data are also shown.
 The dashed line indicates a Schechter function with 
 20\% number density of LBGs at $z \sim 3$.
 Assumed cosmology is $\Omega_\mathrm{M} = 0.3$ 
 and $\Omega_\Lambda =0.7$.
 }
  \label{fig_lf2}
\end{figure}

\citet{S99} and \citet{Adelbarger(2000)} 
made the Schechter function fitting to the luminosity function
of LBGs at $z \sim 3$ and derived a set of parameters 
$m^\ast = 24.54$, $\alpha = -1.57$ and $\phi^\ast = 4.4 \times 10^{-3}$ 
h$^3$ mag$^{-1}$ Mpc$^{-3}$ under the assumption 
$\Omega_M = 0.3$ and $\Omega_\Lambda = 0.7$. 
We made an error-weighted Schechter function fitting 
for our data for $z \sim 5$, and derived 
$m^\ast = 24.5$, $\alpha = -1.50$ and $\phi^\ast = 1.1 \times 10^{-3}$ 
h$^3$ mag$^{-1}$ Mpc$^{-3}$.
In figure \ref{fig_lf2} we show these two Schechter functions 
overplotted on the observed data points.

\begin{figure}
  \begin{center}
    \FigureFile(80mm,50mm){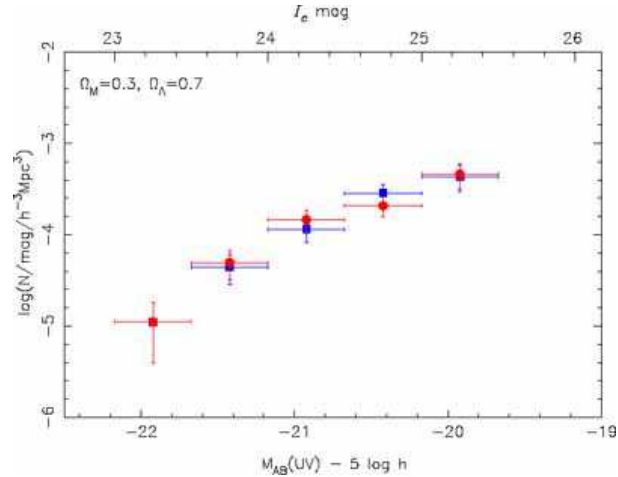}
  \end{center}
 \caption{Luminosity functions of LBGs at $z \sim 5$ with red UV colors
 ($I_c - z' \geq 0.21$, red points) and blue colors ($I_c - z' < 0.21$,
 blue points).
 }
 \label{fig_lf3}
\end{figure}

\citet{Meurer(1999)} suggest that the existence of the correlation 
between the rest-frame UV color and luminosity of the LBGs at 
$z \sim 3$; bright LBGs tend to be red.
\citet{Ouchi01b} report that the number density of LBGs at $z \sim 4$ 
shows the excess of red galaxies,  which is qualitatively consistent 
with the result by \citet{Meurer(1999)}.
In order to see whether such correlation between the UV color and the 
luminosity exists in our sample, we divide it into two subsamples 
by their $I_c - z'$ colors 
(rest-frame 1340\AA{} $-$ 1520\AA{})
at the median value of $0.21$ mag, 
and show luminosity functions of them in figure \ref{fig_lf3}.
In the bright part, the number density of the red galaxies
($I_c - z' \geq 0.21$) is slightly larger than that of the blue
galaxies.
This trend can also be seen in figure \ref{fig_col_mag}.
However, the correlation between the luminosity and the color 
is not as clear as in \citet{Ouchi01b}, who performed the same 
procedure for their LBG candidates at $z \sim 4$ and found 
that red objects are more than two times abundant in the 
brighter part of the luminosity function.
For objects with $I_c \leq 24.5$ in our sample, 
the number of red galaxies are $\sim 1.14$ times larger than 
that of blue galaxies.
Although these results do not conflict with the results 
by \citet{Meurer(1999)} and \citet{Ouchi01b}, the correlation 
of the UV color and luminosity in our $z \sim 5$ sample seems to be 
weaker than those at $z \sim 3$ and $z \sim 4$.

\section{Discussion}

\subsection{Star Formation History of an Individual LBG}

Our results show that the luminosity function of LBGs at $z\sim5$ is
similar to those at $z\sim3$ and at $z\sim4$, though there
may be a slight evolution in the faint part.
Does this imply that the LBGs at $z\sim5$ are the same LBGs  
at $z\sim3$?  Are they continuously forming stars during the
period from $z\sim5$ to $z\sim3$ ($\sim 1$ Gyr in our adopted
$\Lambda$ cosmology)?
\citet{Shapley(2001)} derived the ages of LBGs at $z\sim3$ by
the model SED fitting method assuming constant star formation.
They found that the median age is 320 Myr and that many have 
ages less than 200 Myr and 
only 20\% of them have ages larger than 1 Gyr.
The fraction of the 20\% depends slightly on the luminosity, 
but there is no clear tendency such as brighter objects have 
larger ages or vice versa.
Therefore most of the LBGs at $z\sim3$ were not yet forming
stars at $z\sim5$ and are not direct descendants of
LBGs at $z\sim5$, if we suppose the continuous star formation
history for an individual object.
In figure \ref{fig_lf2} 
we illustrate this with a Schechter function for 
20\% number density of LBGs at $z \sim 3$ as a dashed line, 
by simply assuming that each object has the continuous and 
constant star formation history with an age derived by 
\citet{Shapley(2001)}.
If we compare the number densities 
in the bright part (say $M_{\mathrm{UV}} - 5 \mathrm{log} h \lesssim -21$), 
the number density 
for LBGs at $z \sim 5$ is close to that at $z \sim 3$ and 
the difference is distinct from the 20\% line.
The same situation is depicted by \citet{Ferguson(2002)}.
They also argue that an amount of star formation density at $z\sim5$
contributed by the LBGs seen at $z\sim3$ must be about ten times smaller than
that at $z\sim3$ for the continuous star formation history.

If the LBGs at $z\sim3$ are not the direct descendants of LBGs at
$z\sim5$, did the LBGs at $z\sim5$ stop star formation by
$z\sim3$?  Do they fade out at $z\sim3$?
\citet{Papovich(2001)} point out the absence of massive, 
non-star-forming galaxies in the HDF-N at $z \sim 3$. 
It may indicate that the descendants of LBGs at higher redshift 
which have stopped star formation 
and become fainter in UV by $z \sim 3$ are rare.
Alternatively, the assumption of the constant star formation may not be
appropriate; star formation may be sporadic.
In this case, the ages derived by the SED fitting method would be 
taken to be those of the most recent starburst.
\citet{Papovich(2001)} test SED fitting using two-component models 
comprised of old and young stellar populations, and claim that 
two-component fitting generally gives smaller ages for the
young population than the single-age models.
The sporadic star formation and the short duty-cycle of the starburst
would naturally explain the small ages of young stellar population and 
the existence of older stellar component as well as the 
similarity of the UV luminosity functions at 
$3 \lesssim z \lesssim 5$.
Supposing the sporadic star formation history for individual 
galaxy seems to be reasonable, because 
the hierarchical clustering model generally implies merging 
process must occur for LBGs at high redshift.
In this case, a burst of star formation may occur during a
merging or a galaxy interaction.
In fact, \citet{Nagamine(2002)} showed such sporadic star formation
history for each galaxy takes place associated with a merging
in a large-scale hydrodynamic simulation.
He also presents a resultant rest-frame $V$-band luminosity 
function at $z=5$. We compare it with our result in the UV,
assuming a $UV-V$ color of the model template spectrum with
a moderate reddening.
Although the model luminosity function gives a good agreement 
with the number density of LBGs at the bright part, it shows 
a significant excess ($\sim$1 dex) in the fainter part.
However, a color could depend on luminosity and this 
comparison seems to be premature; we need to compare our result
with a model luminosity function directly derived at UV.

\subsection{Evolution of the UV Luminosity Density}

We can calculate the UV luminosity density at $z \sim 5$ 
by integrating the obtained luminosity function.
In this calculation the integration is made in the magnitude
ranges of $23.0 < I_c \leq 25.0$
and $23.0 < I_c \leq 25.5$.
The $I_c$ magnitudes of 23.0 and 25.5 approximately correspond to 
$M_{\mathrm{UV}}-5 \mathrm{log} h$ of $-22.5$ and $-20.0$
in $\Lambda$-cosmology 
and $-21.7$ and $-19.2$ in Einstein-de Sitter 
cosmology, respectively.
The former integration range is chosen to match with the 
limiting magnitude of the $z \sim 4$ sample by \citet{S99} 
in terms of absolute magnitude.
Results are listed in table \ref{tbl_uvdensity}, together with those 
of $z \sim 3$ and $z \sim 4$ by \citet{S99}.
As it is expected from the shape of the luminosity function
in particular at faint part, 
the resultant UV luminosity density at $z\sim5$ is
0.56 -- 0.69 times of that at $z\sim3$ in the same 
absolute magnitude range, depending on the choice 
of cosmology and the integration range.
The UV luminosity density seems to be getting smaller at higher
redshift, though the difference between $z \sim 5$ and $z \sim 3$-- 4 
is comparable to the uncertainty and we cannot rule out the possibility 
that the UV luminosities are almost constant at this redshift range.
Madau et al. (1996, 1998) derived UV luminosity densities 
at $2 \lesssim z \lesssim 4$ using LBGs in the HDF-N.
Their values are 
smaller than those by \citet{S99} by a factor of 2--4.
The search field for LBGs by Madau et al. (1996, 1998) 
is much smaller than
that by \citet{S99}, and the discrepancy may originate
in the small search volume in the HDF-N due to cosmic variance
\citep{S99}. \citet{Casertano(2000)} also found 1.3 -- 1.9 times 
higher UV luminosity density at $2 \lesssim z \lesssim 4$ 
in the HST/WFPC2 image of HDF-S as compared with that in the 
HDF-N.

To examine a degree of uncertainty in deriving the UV luminosity density, 
we made a calculation of the luminosity function and the UV luminosity
density using the smaller numbers of interlopers (the first test
described in section 5.1).
The resulting UV luminosity density is $\sim$ 1.2 times larger than 
those listed in table \ref{tbl_uvdensity} (an increase of 0.08 in log
scale).
In addition, there is also an uncertainty in the estimate of the
effective volume. 
From the chip-to-chip variation, the uncertainty of the 
effective volume for the whole
survey area is estimated to be $\sim 10$\%.
Taking these two uncertainties into account, 
the upper most possible UV luminosity density is estimated to be 
about 1.6 times ($\sim 0.2$ dex) larger
than that shown in table \ref{tbl_uvdensity}.
The  value is comparable to the luminosity densities at 
$z \sim 3$ listed in table \ref{tbl_uvdensity}.

\begin{figure}
  \begin{center}
    \FigureFile(80mm,50mm){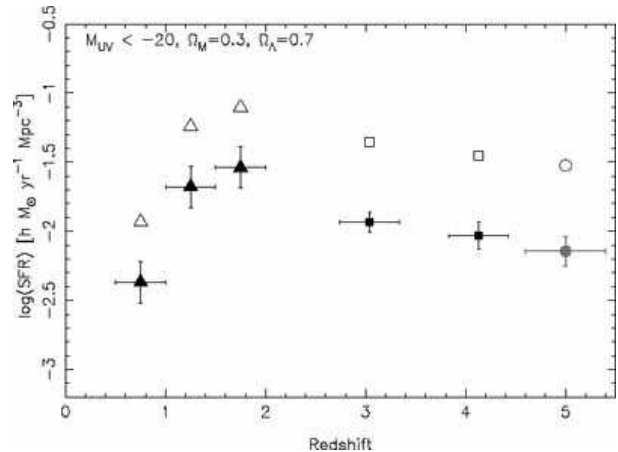}
  \end{center}
 \caption{
 The star formation density as a function of redshift.
 Our data point for $z \sim 5$ LBGs are marked as a red filled
 circle. Our data and those by \citet{S99} at $z \sim 3$ and 4 
 are derived by the integration of the luminosity function 
 with absolute UV magnitude $M_{\mathrm{UV}}-$ 5 log $h$ $\lesssim -20.0$ 
 (see table \ref{tbl_uvdensity}).
 Three data points (shown in filled triangles) 
 for $z \sim 0.5$ to 2.0 is from 
 \citet{Connolly(1997)}, which are adjusted to 
 $\Omega_{\mathrm{M}}=0.3$, $\Omega_\Lambda=0.7$ cosmology and 
 to represent UV luminosities from objects with 
 $M_{\mathrm{UV}}-$ 5 log $h$ $\leq -20$.
 The conversion factors from UV luminosity (at rest-frame 1500\AA 
 for $z \sim 3$ to 5 and 2800\AA for $z < 2$) to star formation density
 is from \citet{Madau(1998)}.
 Filled symbols indicate values without correction for dust extinction.
 Dust extinction is corrected following the 
 prescription by \citet{Calzetti(2000)} with $E(B-V) = 0.15$ for 
 all data points, and the values are shown with open symbols.
 Assumed cosmology is $\Omega_{\mathrm{M}}$ = 0.3,
 $\Omega_\Lambda$ = 0.7.
 } 
  \label{fig_sfh1}
\end{figure}

In figure \ref{fig_sfh1} we show the star formation density as a function
of redshift. The assumed cosmology is $\Omega_\mathrm{M} = 0.3$ and
$\Omega_\Lambda = 0.7$. 
The data point at $z \sim 5$ from our data and the points at 
$z \sim 3$ and 4 from \citet{S99} are derived by the integration of 
the observed number densities for objects down to
$M_{\mathrm{UV}} - 5 \mathrm{log} h\lesssim -20$, 
as listed in table \ref{tbl_uvdensity}.
For these data points at $z \gtrsim 3$, the UV luminosity density at 
1500 \AA\ is converted to star formation density 
using a conversion factor by \citet{Madau(1998)}, i.e., 
{\it SFR} ($\MO$ yr$^{-1}$) = $L_{\mathrm{UV}}$ (erg s$^{-1}$
Hz$^{-1}$) $\times 1.25 \times 10^{-28}$.
For $z \sim 5$ data point this gives {\it SFR} of 
$7.18\times10^{-3}$ h $\MO$ yr$^{-1}$ Mpc$^{-3}$ 
($-2.14$ in log scale) and $1.30\times10^{-2}$ h $\MO$ yr$^{-1}$ Mpc$^{-3}$ 
($-1.88$ in log scale) for $\Lambda$ cosmology and Einstein-de
Sitter cosmology, respectively.
The data points for $z \sim 0.5$ -- 2.0 are taken from
\citet{Connolly(1997)} which is based on the UV luminosity densities 
at rest-frame 2800 \AA. For this wavelength we adopt 
a conversion factor 
{\it SFR} ($\MO$ yr$^{-1}$) = $L_{\mathrm{UV}}$ (erg s$^{-1}$
Hz$^{-1}$) $\times 1.27 \times 10^{-28}$ 
also given by \citet{Madau(1998)}. 
Originally their UV luminosity densities are
given by integrating the Schechter UV luminosity function  
over the all luminosity range and calculation is made for 
Einstein-de Sitter universe.
We adopt here a Schechter function with $\alpha = -1.3$, and 
for the consistency with data at $z \geq 3$, the integration of the 
UV luminosity function is made down to 
$M_{\mathrm{UV}} - 5 \mathrm{log} h = -20$.
The correction for different cosmological parameters is made following
the prescription by \citet{Somerville(2001)}.
In figure \ref{fig_sfh1} we also indicate the star formation density 
with a simple dust extinction correction as open symbols. 
In correcting dust extinction we follow \citet{S99}, 
who use the extinction curve by \citet{Calzetti(2000)} and assume 
$E(B-V)=0.15$ for all data points. As \citet{S99} remark, the validity
of such uniform correction is not clear, since the degree of extinction 
may be redshift dependent, and dust properties may change form time to
time. As it is described in section 3.4, 
the $E(B-V)$ distribution estimated from $I_c - z'$ colors 
at $z \sim 5$ is similar to that of galaxies at $z \sim 3$ 
(figure \ref{fig_color}), although the photometric error is large for 
faint objects. Thus the assumption that the degree of dust extinction 
is similar from $z \sim 3$ to 5 might not be
inconsistent with our observed results for LBGs at $z \sim 5$, 
although there may be heavier dust extinction at $z \sim 1$ 
\citep{Tresse(2002)}.

We should note that the values of star formation density appeared 
in figure \ref{fig_sfh1} are smaller than those in figure 9 of
\citet{S99}. This is because our integration limit of the luminosity 
function is about 0.5 mag brighter than that of \citet{S99}.
The position of data points for $z \sim 1.75$ 
is about 0.4 dex higher relative to $z \sim 3$ in figure \ref{fig_sfh1},
while in \citet{S99} the difference is $\sim 0.2$ dex.
This is caused by the difference of the adopted slope of the fainter 
part of the luminosity function between $z < 2$ and $z =3 \sim 4$.

\begin{table}
\begin{center}
\caption{UV luminosity densities.}
 \begin{tabular}{llllll}
  \hline\hline
  &\multicolumn{4}{c}{log $\rho_{\mathrm{UV}}$ [h erg s$^{-1}$ Hz$^{-1}$ Mpc$^{-3}$] [$\ast$]}\\
  \cline{2-6}
  &\multicolumn{2}{c}{$\Omega_{\mathrm{M}}$ = 0.3, $\Omega_\Lambda$ = 0.7} & & 
   \multicolumn{2}{c}{$\Omega_{\mathrm{M}}$ = 1.0, $\Omega_\Lambda$ = 0.0}\\
  \cline{2-3} \cline{5-6}
  redshift & $M_{\mathrm{UV}}-$5 log $h$ $\lesssim -20.5$ & 
   $M_{\mathrm{UV}}-$ 5 log $h$ $\lesssim -20.0$ & & 
   $M_{\mathrm{UV}}-$ 5 log $h$ $\lesssim -19.7$ & 
   $M_{\mathrm{UV}}-$ 5 log $h$ $\lesssim -19.2$\\ 
  \hline
  $z = 3$[$\dagger$]  & 25.75$\pm$0.07        & 25.97$\pm0.07$ &    & 26.05$\pm$0.07 & 26.27$\pm 0.07$ \\
  $z = 4$[$\dagger$]  & 25.70$\pm$0.10 (89\%) & --    &    & 25.96$\pm$0.10 (81\%)& --\\
  $z = 5$       & 25.59$\pm$0.11 (69\%) & 25.76$\pm$0.16 (62\%)& & 25.85$\pm$0.11 (63\%)& 26.02$\pm$0.16 (56\%)\\
  \hline

  \multicolumn{4}{@{}l@{}}{\hbox to 0pt{\parbox{120mm}{\footnotesize
  \par\noindent
  \footnotemark[$\ast$] UV luminosity density is obtained directly from the integration of 
  the observed number densities.
  Values shown in parentheses are percentage of
  the UV luminosity density relative to that at $z \sim 3$.
  \par\noindent
  \footnotemark[$\dagger$] \citet{S99}.
  }\hss}}
 \end{tabular}

 \label{tbl_uvdensity}
\end{center}
\end{table}

The overall shape of the star formation history 
presented in 
figure \ref{fig_sfh1} shows a steep rise from 
$z = 0$ to 2, and appears to be almost constant or slowly 
declining at larger redshift.
This tendency has been 
reproduced by recent hydrodynamical simulations 
(e.g., \cite{Nagamine(2001)}; \cite{Ascasibar(2002)}) 
and semi-analytic models of galaxy formation which 
take collisional starburst into account 
(\cite{Somerville(2001)}; \cite{Balland(2002)}).
It is hard to make a direct comparison of the observed 
data and these predictions, because the observations 
(especially for high-redshift objects) are restricted 
to trace only the bright part of the luminosity 
function. 
Although it would be desirable if the model predictions with 
an integration range comparable to the observational 
limit are available, we do not know such 
predictions among the recently published semi-analytic 
models and numerical simulations; they usually give 
the total star formation density.
The shape of the faint end of the luminosity function 
has a crucial effect on the UV luminosity density and 
may differ according to the redshift. 
It is also needed to consider the uncertainty of dust 
extinction in the UV luminosity, 
the correction factor often amounts to more than 3.
We might be able to say, however, that 
if we assume that the shape of the UV luminosity function 
and the degree of dust extinction 
do not change significantly with respect to the redshift, 
recent semi-analytic models with collisional starburst 
such as by \citet{Somerville(2001)} and hydrodynamical 
simulations such as by \citet{Ascasibar(2002)} are 
broadly consistent with the observed data including our 
result for $z \sim 5$.

\subsection{Contribution of Stellar Sources to the Ionizing Photon 
Density in the Intergalactic Medium}

Finally we briefly comment on the contribution of stellar sources 
to the hydrogen-ionizing radiation field at $z \sim 5$. 
Recently \citet{Steidel(2001)} claimed that the escape fraction 
$f_{\mathrm{esc}}$, 
which represents the rate of Lyman continuum photons emitted outside
from a galaxy, of the LBGs at $z \sim 3$ is significantly 
higher than the observational upper limit for star-forming galaxies 
in the local universe; 
it is comparable to or larger than 0.5
(but \citet{Giallongo(2002)} made a deep spectroscopic 
observation for two LBGs and placed upper limit of 
$f_{\mathrm{esc}} < 0.16$).
\citet{Steidel(2001)} estimated the contribution of LBGs at $z \sim 3$ 
to the ionizing radiation field at 1 ryd 
to be $(1.2 \pm 0.3) \times 10^{26}$ 
h erg s$^{-1}$ Hz$^{-1}$ Mpc$^{-3}$, which is 
about 5 times larger than the contribution from quasars
at the same redshift \citep{Madau(1999)}.
If the rapid decrease of quasar number density at $z > 3$ 
(\cite{Fan(2001)} and references therein)
is real, the number of ionizing photons radiated from quasars 
should strongly decline at higher redshift.
\citet{Madau(1999)} estimated it drops by a factor of $\sim$2 
from $z=3$ to $z=5$.
Provided that the escape 
fraction of LBGs at $z \sim 5$ is as large as 0.5, 
our estimate of the UV luminosity density at $z \sim 5$ 
implies the ionizing photons originated from star forming 
galaxies are $\sim$5 to 7 times larger than those from quasars 
at the redshift.
This further suggests that the source of the UV photons which reionized the 
Universe at sometime of $z > 5$ is dominated by stellar origin.
It is also interesting that, the combination of the estimated numbers of 
the ionizing photons from LBGs and quasars at $z \sim 5$ is comparable to the 
number of photons needed to fully ionize a universe considered 
in \citet{Madau(1999)} (Einstein-de Sitter universe 
with a clumping factor of 30 and the baryon 
fraction $\Omega_b$h$^2$=0.02).
However, this coincidence is based on the quite uncertain factors 
such as the quasar number density, the escape fraction of ionizing photons, 
and the clumpiness of the hydrogen gas in the early universe.
A deep observation of the rest-frame far-UV wavelengths of 
LBGs is one of the indispensable subjects needed to clarify the nature of 
the ionizing radiation field at $z > 3$.

\section{Summary and Conclusion}

We have presented the results of a search for Lyman break galaxies 
at the redshift range between 4.4 and 5.3 
in the field including the HDF-North 
with Suprime-Cam attached to Subaru telescope.
Databases for galaxies in the HDF-N and its flanking fields 
enable us to find out the color selection criteria 
with least foreground contamination.
Within the surveyed area of 618.6 arcmin$^2$ 
we have found 96 LBG candidates with $23.0 < I_c  \leq 24.5$ mag, 
and 310 candidates with $I_c \leq 25.5$ mag.
There is a hint of the deficiency of bright blue galaxies, 
although it is not as clear as suggested for LBGs at $z \sim$ 3 to 4. 

The contamination of objects at intermediate redshifts 
is estimated employing the resampling algorithm. The fraction of 
contamination amounts to $\sim$20\% at bright part of the sample 
and $\sim$50\% at the faintest magnitude range.
We have also performed Monte Carlo simulations for the evaluation of the 
incompleteness of the survey. The completeness is estimated 
as a function of the magnitude and the redshift of an object; 
in the redshift range of $4.5 \lesssim z \lesssim 5.0$, 
it is $\sim$80\% in the bright part of the sample and 
$\sim$20\% in the faintest magnitude range.
Using these corrections, we derive the 
luminosity function of LBGs at $z \sim 5$ statistically in the 
absolute UV magnitude range from $-23.0$ to $-20.5$ 
(with $\Omega_{\mathrm{M}} = 0.3$, $\Omega_\Lambda = 0.7$, $h_0 = 0.65$).
No significant difference is seen between the luminosity function of 
LBGs at $z \sim 5$ and those at $z \sim 3$ and 4 obtained by 
\citet{S99}, 
though there might be a decrease
of the number density at the fainter magnitude.
We have compared the UV luminosity density at the redshift 5 
to those at $z \sim 3$ and 4.
If we adopt the direct sum of the luminosity function within 
our observational limit, the star formation density 
decreases to 56--69\% of that at $z \sim 3$ estimated by \citet{S99}.
This decrease in UV luminosity density at $z \sim 5$ compared 
to $z \sim 3$ is due to the smaller number density of faint galaxies at 
$z \sim 5$. We cannot rule out the possibility that the UV luminosity 
density is almost constant at $z \gtrsim 3$, if we take account for the 
larger uncertainty in the faint part of the luminosity function.

We have discussed on a plausible evolutionary scenario of LBGs based on 
our results. 
The similarity of the luminosity functions at redshifts 5 to 3 
implies that most of the LBGs at $z \sim 5$ should have faded out 
at $z \sim 3$ and the LBGs at $z \sim 5$ are different galaxies 
from those seen at $z \sim 3$, 
if we take face values for ages of the LBGs at $z \sim 3$ obtained 
by the SED fitting in which a continuous star formation in an 
individual galaxy is assumed.
However, if the star formation in LBGs is sporadic, the similarity 
of the luminosity function at $z \sim 3$ and 5 would be
 naturally explained. 
Such sporadic star formation has been suggested by hydrodynamical 
simulations and semi-analytic models with collisional starbursts, 
and the trend of the cosmic star formation history 
predicted by these studies resembles to 
that estimated from an integration of the UV luminosity 
density within the observational limit.

\bigskip

We thank all staff of Subaru telescope during our observations, 
especially Yutaka Komiyama, for their devoted support and 
technical assistance. We also appreciate Suprime-Cam group 
members, especially Masami Ouchi, 
for their kind advices and discussions 
for data reduction procedures, and the anonymous 
referee for giving us insightful suggestions.
II is supported by a Research 
Fellowship of the Japan Society for the Promotion 
of Science for Young Scientists.

\end{document}